\documentclass[12pt]{article}
\usepackage{amssymb} 
\def\href#1#2{#2}
\def\IP{\relax{\rm I\kern-.18em P}}

\newcommand{\nc}{\newcommand}
\newcommand{\beq}{\begin{equation}}
\newcommand{\eeq}{\end{equation}}
\newcommand{\beqa}{\begin{eqnarray}}
\newcommand{\eeqa}{\end{eqnarray}}
\def\tQ{{\hat Q}} 
\def\bj{{\bar \jmath}} 
\def\cA{{\cal A}}
\def\cV{{\cal V}}
\def\cH{{\cal H}}
\def\cJ{{\cal J}} 

\nc{\nn}{\nonumber}
\def\e{\epsilon}
\def\o{\otimes}

\def\IP{\relax{\rm I\kern-.18em P}}
\def\tr{{\rm tr\ }}
\def\mod{{\rm mod}}
\def\CN{{\cal N}}
\def\QN{\mathbb{N}}

\def\QR{\mathbb{R}}

\def\QZ{\mathbb{Z}}
\def\QP{\mathbb{P}}
\def\U{{U}}

\def\kket#1{|#1\rangle\rangle}
\def\vev#1{\langle{#1}\rangle}
\def\ppe{\hspace*{-2.5mm}}
\def\ew{\hspace*{-1mm}}
\newcommand{\Fus}[6]{F_{{\scriptstyle #1}{\scriptstyle #2}}
  \hspace*{.3mm}\displaystyle{[} \ew \begin{array}{ll} {\scriptstyle #3 }
  \ppe & {\scriptstyle #4} \ppe \\[-2mm] {\scriptstyle #5}\ppe &
  {\scriptstyle #6}\ew \end{array}\displaystyle{]}}
\newcommand{\Br}[6]{B_{{\scriptstyle #1}{\scriptstyle #2}}
  \hspace*{.3mm}\displaystyle{[} \ew \begin{array}{ll} {\scriptstyle #3 }
  \ppe & {\scriptstyle #4} \ppe \\[-2mm] {\scriptstyle #5}\ppe &
  {\scriptstyle #6}\ew \end{array}\displaystyle{]}}
\newcommand{\CG}[6]{\displaystyle{[} \,\ew \begin{array}{lll} 
  {\scriptstyle #1} \ppe
  & {\scriptstyle #2} \ppe & {\scriptstyle #3} \ew \\[-2mm] {\scriptstyle
  #4} \ppe & {\scriptstyle #5}\ppe & {\scriptstyle #6} \ew\end{array}
  \displaystyle{]}}

\newcommand{\norm}[9]{\ew \begin{array}{lllll}
  | & \ppe {\scriptstyle #1} & \ppe {\scriptstyle #2} & 
    \ppe {\scriptstyle #3} & \ppe \ew | \\[-2mm] 
  | & \ppe {\scriptstyle #4} & \ppe {\scriptstyle #5} & 
    \ppe {\scriptstyle #6} &\ppe \ew | \\[-2mm] 
  | & \ppe {\scriptstyle #7} & \ppe {\scriptstyle #8} & 
    \ppe {\scriptstyle #9} &\ppe \ew | \end{array}\ppe}    
\newcommand{\inorm}[9]{\ew \begin{array}{lllll}
  | & \ppe {\scriptstyle #1} & \ppe {\scriptstyle #2} & 
    \ppe {\scriptstyle #3} & \ppe \ew |^{-1} \\[-2mm] 
  | & \ppe {\scriptstyle #4} & \ppe {\scriptstyle #5} & 
    \ppe {\scriptstyle #6} &\ppe \ew | \\[-2mm] 
  | & \ppe {\scriptstyle #7} & \ppe {\scriptstyle #8} & 
    \ppe {\scriptstyle #9} &\ppe \ew | \end{array}\ppe}    
\def\orb{{\mbox{\rm \scriptsize orb}}}
\def\nn{\nonumber}
\def\hp{{\hat{+}}}
\def\cH{{\cal H}}
\def\BZ{\QZ}
\def\S{{\cal S}}
\def\proj{{\mbox{\rm \scriptsize proj}}}
\def\a{\alpha}
\def\ti{\times}
\def\smathbfn#1{{\mbox{\boldmath $\scriptstyle #1 $}}}
\def\mathbfn#1{{\mbox{\boldmath $ #1 $}}}
\newcommand{\vvert}[3]{\displaystyle{(} \ew \begin{array}{ll} 
  \ & \hspace*{-4mm} {\scriptstyle #1} \\[-2mm] {\scriptstyle #2} 
  \ppe & {\scriptstyle #3} \ew \end{array} \displaystyle{)}}      
\renewcommand{\vert}[3]{\displaystyle{(} \ew \begin{array}{ll} 
  \ & \hspace*{-2mm} {\scriptstyle #1} \\[-2mm] {\scriptstyle #2} 
  \ppe & {\scriptstyle #3} \ew \end{array} \displaystyle{)}}      
\newcommand{\su}{{\widehat{SU}(2)_k}}
\input epsf
\newcount\figno
\def\fig#1#2#3{
\par\begingroup\parindent=0pt\leftskip=1cm\rightskip=1cm\parindent=0pt
\baselineskip=11pt
\global\advance\figno by 1
\epsfxsize=#3
\centerline{\epsfbox{#2}}
\vskip 12pt
{\bf Figure \the\figno:} #1\par
\endgroup\par
}
\def\figlabel#1{\xdef#1{\the\figno 
\mbox{ }}}
\def\encadremath#1{\vbox{\hrule\hbox{\vrule\kern8pt\vbox{\kern8pt
\hbox{$\displaystyle #1$}\kern8pt}
\kern8pt\vrule}\hrule}}

\begin{document}
\baselineskip=17pt
\title{\bf On Superpotentials for D-Branes \\[2mm] 
 in Gepner Models  \\[5mm] }
\author{{\sc Ilka Brunner} \\[2mm] 
 Dept.\ of Physics and Astronomy, Rutgers University, \\
 Piscataway, NJ 08855, U.S.A.  \\[5mm]  
     {\sc   Volker Schomerus} \\[2mm]
Albert-Einstein-Institut
\\ Am M\"uhlenberg 1, D--14476 Golm, Germany \\[1mm] }
\vskip.2cm
\date{August 24, 2000}
%
\begin{titlepage}      \maketitle       \thispagestyle{empty}

\vskip1cm
\begin{abstract}
\noindent  
A large class of D-branes in Calabi-Yau spaces can be constructed 
at the Gepner points using the techniques of boundary conformal 
field theory. In this note we develop methods that allow to compute 
open string amplitudes for such D-branes. In particular, we present 
explicit formulas for the products of open string vertex operators
of untwisted A-type branes. As an application we show that the 
boundary theories of the quintic associated with the special 
Lagrangian submanifolds $ \Im \omega_i z_i = 0$ where 
$\omega_i^5=1$ possess no continuous moduli.         
\end{abstract}
\vspace*{-18.9cm}
{\tt {AEI-2000-050 \hfill RUNHETC 2000-32 }}\\
{\tt \phantom{{UUITP-nn/99} \hfill hep-th/0008194}}
\bigskip\vfill
\noindent
\phantom{wwwx}{\small e-mail:}{\small\tt ibrunner@physics.Rutgers.EDU; 
vschomer@aei-potsdam.mpg.de} 
\end{titlepage}

\section{Introduction}

It has been known for a long time \cite{wittph, Gepn1, Gepn2} that 
certain rational $\CN=2$ conformal field theories, the so-called 
Gepner models,  can be used to describe closed strings in the 
small volume regime of geometrical Calabi-Yau compactifications. 
More recently, it was attempted to extend this correspondence to
the open string sector. This discussion was initiated in \cite{BDLR} 
through a comparison of D-branes on the quintic to the rational 
boundary theories for the Gepner model $(k=3)^5$ obtained in \cite{RS}. 
The results were then extended to other models in \cite{DiaRom,KLLW,ej,
Scheid,DFR2}. Significant progress has been made in matching  
B-type boundary states with the brane spectra at large volume 
\cite{DFR1,DFR2,DouDia}. Related work using  Landau-Ginzburg 
and gauged linear sigma model methods
can be found in \cite{HIV,suresh1,suresh2,GJSIII}.
\smallskip

Until now, most of the conformal field theory results deal with the 
construction of boundary states at Gepner points \cite{RS,NakNoz,BruSch,
FSW,suresh1}. These states allow to compute the open string partition 
functions and thereby to determine the spectrum of operators in the 
small volume limit. However, we can only get insight into the world-%
volume theory on the brane if we know how the world-volume fields 
interact. This information is encoded in operator product expansions
(or correlation functions) of vertex operators at the boundary of the 
2-dimensional world-sheet. The aim of the present paper is to provide 
the tools that are needed to compute correlation functions of open
string vertex operators for a large set of rational branes in Gepner 
models. We are going to concentrate on A-type boundary states, but our  
method can be extended to the B-type situation as well.
\smallskip

Superpotentials for the massless fields are of particular interest 
because they determine the moduli spaces of branes. There exists a 
powerful theorem due to MacLean, which states that at the large 
volume point, all moduli of A-type branes are unrestricted. 
When rephrased in the framework of CFT the theorem claims that all 
marginal boundary operators are in fact truly marginal. According to 
\cite{KKLM1,KKLM2}, the theorem breaks down once we are forced to take 
world-sheet instantons into account. Hence, when we pass to smaller
volumes we expect to find restricted moduli. Clearly, this is to 
be expected from a general world-sheet point of view \cite{RS2}. 
Below we shall demonstrate such a restriction through an explicit 
example. In fact, there exist boundary theories for the quintic 
supporting a massless field at the Gepner point which is not present 
at large volume \cite{BDLR}. We shall apply our general techniques to 
these boundary theories and show that there is no continuous modulus 
associated with this massless mode.
\smallskip

Beyond the investigation of truly marginal operators and continuous 
moduli, computations of correlation functions are also essential for
the analysis of unstable brane configurations and the identification 
of their bound states. Decays caused by tachyon condensation have 
been discussed a lot in the recent literature (see e.g. \cite{Sen,
HKM}). There exist other scenarios where gauge fields on a stack of 
branes condense \cite{ARS}. Our understanding of such condensation 
phenomena in non-trivial compactifications is still very limited
(but see \cite{Sen2,DFR2,DouDia} for some results in this
direction).  
\medskip

This paper is organized as follows: We are starting out with a 
brief review of the general structure of open string vertex operators 
and their correlation functions in $\CN=2$ superconformal field 
theories (Section 2). Then we turn to a more detailed analysis of 
boundary conformal field theories describing the internal sector of 
a Gepner compactification. The first step is performed in Section 3
and it involves some general results dealing with correlation functions
of open string vertex operators on a particular class of orbifolds. 
Section 4 contains a more technical computation of the fusing matrix
of $\CN=2$ minimal models. The results of Section 4 are then combined 
with the general analysis from Section 3 to determine the operator 
product expansion of open string vertex operators for A-type branes
in Gepner models. After a brief background review on branes in 
Gepner models we will present the main formulas of this text in 
Section 5.3. Finally, in Section 6, we shall apply our formalism 
to certain boundary states of the quintic. The aim here is to 
show that the moduli spaces at large and small radii do agree
after superpotential terms have been taken into account. Note 
that this is not guaranteed by the decoupling conjecture of 
\cite{BDLR} because the latter only makes statements about the
(in)dependence of B-type moduli spaces on the K\"ahler modulus.

\section{Open string correlators in $\CN=1$ theories.}

Let us consider a BPS brane placed in a $\CN=2$ compactification 
of type II string theory. It is well known that this scenario leads
to a 4-dimensional $\CN=1$ supersymmetric field theory on the 
world-volume of the brane, provided the brane extends in all the 
non-compact directions. The effective field theories may contain 
a number of massless chiral and vector superfields. The former 
are composed from a scalar $\phi$, a fermionic component $\psi$ 
and an auxiliary field $F$. Vector superfields consist of a vector 
$v$, a fermionic field $\lambda$ and an auxiliary component $D$. 
In principle, both $D$- and $F$-terms can contribute to the 
potential for scalar fields. Here we are interested in computing 
superpotentials for the chiral world volume fields. Hence, the 
leading non-linear term in the corresponding superpotential is of 
the form $\phi \psi \psi$ or, equivalently, $F \phi \phi$.  
\medskip

To compute these terms from conformal field theory, we have to assign 
vertex operators to all the fields we are interested in. In particular, 
the world-volume scalars $\phi^a$ \footnote{We shall use the superscript 
$a$ to distinguish between different massless chiral superfields} are 
represented by vertex operators in the Neveu-Schwarz (NS) sector. 
Within the $(-1)$ picture, they are given by
$$
\cV_{\phi^a}^{(-1)}(x) \ = \ e^{-\Phi(x)}\, \psi^a_\phi(x)\ \ . 
$$
Here, $\Phi$ denotes the bosonized superconformal ghost and 
$\psi^a_\phi$ is a NS operator of the internal theory. Both 
factors have dimension 1/2 in the case that the scalar is massless. 

For world-volume fermions one uses Ramond vertex operators. When 
we write them in their canonical $(-1/2)$ picture, they are of the 
form 
$$
\cV_{\psi^a}^{(-1/2)}(x) \ = \ \xi^{a,\alpha}_i \, S_{\alpha}(x) 
   e^{-\frac{1}{2}\Phi(x)} \, \Sigma^{a,i}(x)\ \ ,
$$
where $\xi^{a,\alpha}_i$ describes the polarization, $S^\alpha$ 
is the space-time part of the spin field which has dimension 
1/4. The dimension $3/8$ fields $\Sigma^i(x)$ are associated 
to Ramond ground states of the internal sector.

Vector fields $v$ come with the NS-sector again. Their vertex 
operators have the identity field in the internal part,   
$$
\cV^{(-1)}_{v}(x) \ = \ \xi_\mu \, e^{-\Phi(x)} \, \psi^\mu(x)\ \ .  
$$
$\psi^\mu$ is a vector of world-sheet fermions and $\xi$ describes 
the polarization. One may add Chan-Paton matrices to all three 
vertex operators we have described. After this extension, the 
vector fields can give rise to non-abelian gauge fields. 

The factors $\psi^a, \Sigma^{a,i}$ which the internal part contributes
to the vertex operators, carry the label $a$ depending on the pair 
of boundary conditions we consider for the two ends of open strings. 
We shall be much more specific about these labels later when we turn 
to concrete models. 
\medskip

Before spelling out which correlators we want to compute, we 
would like to construct the space-time supersymmetry generators
$Q_\alpha(x)$. To this end, let us introduce a free bosonic 
field $X$ for the internal $U(1)$ current $J(x)$ of the 
superconformal algebra, 
$$
J(x) \ = \ i \sqrt{\frac{c}{3}} \partial X(x) \ \ ,
$$
$X$ appears together with the space-time spin field 
$S^\alpha(x)$ in the following formula for $Q_\alpha$ 
\beq\label{sf}
 Q_{\alpha}^{(\pm 1/2)}(x) \ = \ 
 e^{\pm \frac{\Phi(x)}{2}} e^{\pm  \frac{i}{2} \sqrt{\frac{c}{3}} X(x)}
 S_{\alpha}(x)
\eeq
The exponential $\exp(\pm i \eta \sqrt{c/3} X(x))$ generates a spectral 
flow by $\eta$ units in the internal part of the theory. Hence, the 
formula (\ref{sf}) for the supersymmetry generators $Q_\alpha$ contains 
the spectral flow operator for a flow by $1/2$ units. Note that the 
application of $Q_a$ to the vertex operator for scalars gives the
vertex operator of a fermionic field because the spectral flow by 
$1/2$ unit maps the NS- into the R-sector. One unit of spectral flow 
(and hence an action of two space-time supersymmetry generators) is 
needed to construct the vertex operator of the auxiliary field $F^a$ 
{}from the vertex operator for the scalar $\phi^a$. 
\medskip

In computing the world-sheet correlators we have to respect two 
simple rules. First the total $\Phi$-charge of fields inside the
correlation function must add up to $-2$. If necessary, we have 
to use the vertex operators in pictures different from the 
canonical ones we spelled out above. The second rule states that 
three of the vertex operators should be multiplied with the 
anti-commuting ghost fields $c$. After this, one has to integrate 
over the world-sheet arguments of the boundary fields and to sum 
over arbitrary permutations, as usual. 
\smallskip

Let us give an explicit formula for the correlator that computes
the leading non-linear contribution $\phi \psi \psi$ to the 
superpotential for the scalar $\phi$. In this case, we have 
to compute a 3--point function and we can use all three vertex 
operators in their canonical pictures since their $\Phi$ charge 
add up to $-2$, 
\beq \label{ppp}
   \phi \psi \psi\ :  \ \ \ \ \ 
   \langle c(x_1) \cV^{(-1)}_\phi(x_1) \ c(x_2) \cV^{(-1/2)}
   _\psi(x_2) \ c(x_3) \cV^{(-1/2)}_\psi(x_3) \rangle  \ + \ 
   2 \leftrightarrow 3 \ \ . 
\eeq
Similarly, one could compute the $F \phi \phi$ term by 
applying one full unit of spectral flow (and of $\Phi$-charge) 
to the vertex operator for scalars. The computation of 
such correlation functions factors into several independent 
parts. Contributions from the ghosts $c$, the space-time 
spin fields $S_\a$ and the superconformal ghosts $\Phi$ are 
evaluated easily since this involves only calculations in a
free field theory. For this reason, we shall focus mainly 
on the correlation functions of the fields $\psi^a, 
\Sigma^{a,i}, X$ in the internal sector.  
\smallskip 

Most Calabi-Yau compactifications are described by an 
interacting internal conformal field theory which makes 
the calculations rather difficult. But we can at least 
illustrate the computations we have in mind through 
one simple example where the internal part is given 
by a free field theory, too. Namely, we consider D3 
branes in a torus compactification. Their world-volume 
theory is an $\CN=4$ gauge theory. In $\CN=1$ language, 
it contains three chiral superfields $Z^a$. In this case, the 
superpotential is known to be of the form $W=\tr Z^1[Z^2, 
Z^3]$. Here the trace is over the Chan-Paton matrices 
which are non-trivial if we stack several D3 branes. The
superpotential $W$ can easily be reproduced by a world-%
sheet computation. To this end, we take three vertex 
operators for world volume scalars:
\beq\label{scalar}
\cV^{(-1)}_{\phi^a}(x) \ = \ e^{-\Phi(x)} \psi^a_\phi(x) \ \ ,
\eeq
where the $\psi^a_\phi= \psi^a$ are the three (fermionic) 
chiral primary 
fields of charge one that exist on a 3 complex dimensional 
torus. According to our general discussion we need to add one
unit of spectral flow to one of the fields in the internal 
theory (or, alternatively, two $1/2$ units to two fields).
In the case at hand, the spectral flow operator is $(\psi^a 
\psi^b \psi^c)^\dagger$. This leads to a three-point function 
that is proportional to the antisymmetric tensor $\epsilon_{abc}$
reproducing the known result. Note that the string amplitude
vanishes for a single brane.%
\smallskip%

The last observation raises the question, whether the allowed 
superpotential terms on a single brane vanish generically after 
summing over all permutations of the insertion points at the 
boundary of the world-sheet. But we learn from the relation 
between open string theory and non-commutative geometry 
\cite{DoHu,ChuHo,Scho,SeiWit} that the products of boundary 
operators in a conformal field theory (such as the Gepner 
models) can depend very much on the order in which fields 
are multiplied. Only a small number of fields, namely those which can be 
analytically continued into the bulk, are protected against 
such effects. This happens, for instance, in case of the free
fermionic fields $\psi^a$ which appeared in our discussion of 
$D3$ branes in a torus compactification. Similar arguments may 
apply to marginal operators built from fermionic fields 
in the large volume limit, but they are expected to break 
down when we get into the small volume regime. The aim of this
paper is to substantiate such expectations by rigorous
conformal field theory computations at the Gepner point. 

\section{Boundary conformal field theory on orbifolds}

In this section we present some basic material relevant to 
the investigation of D-branes on orbifolds. Under certain 
simplifying assumptions on the nature of the orbifold action
and on the D-branes under consideration, we shall present a 
general formula for the operator product expansion of boundary 
operators. 

\subsection{Boundary conformal field theory on the covering space} 
To begin with, let us review the necessary input from 
Cardy's work. Suppose we are given some bulk conformal
field theory with chiral algebra $\cA$ and a modular
invariant partition function of the form 
\beq 
Z(q,\bar q) \ = \ \sum_j \ 
\chi_j(q) \ \chi_{\bj} (\bar q)\ \ \ . \label{Bpart}
\eeq 
Here $j$ runs through the sectors of $\cA$ and $\bj$ is
just another sector that  appears together with $j$ in 
the partition function. $\chi_i(q)$ denotes the character 
that comes associated with the sector $i$ of the chiral 
algebra. 
\smallskip

It was explained in \cite{RS} that the construction 
of boundary theories involves picking some automorphism 
$\Omega$ of the chiral algebra. This appears in the 
boundary conditions to describe how left- and right 
movers are glued along the boundary. Any such automorphism
$\Omega$ induces a map $\omega$ that acts on sectors $i$ 
of the chiral algebra. Cardy's analysis of boundary 
conditions applies whenever $\omega(j)^\vee = \bj$. 
Here $i^\vee$ denotes the sector conjugate to $i$, i.e. 
the unique label with the property that its fusion 
product with $j$ contains the vacuum representation 
$0$ of the chiral algebra. We will call such a modular 
invariant ($\Omega$)-diagonal.%
\smallskip%

Under this condition, Cardy provides us with a list of 
boundary theories. Their number agrees with the number 
of sectors of $\cA$. We will use labels $I,J,K, \dots$ 
to distinguish between boundary conditions and sectors 
but it should be kept in mind that small and capital 
letters run through the same index set. The spectrum 
of open strings that stretch between the branes that 
are associated with the labels $I$ and $J$ is given by 
\beq Z_{IJ}(q) \ = \ \sum_j N_{Ij}^{\ \ J} \ \chi_j(q) 
\ \ . \label{bpart} \eeq
Obviously, this tells us how the state space $\cH_{IJ}$ of 
the boundary theory is built up from sectors of the 
chiral algebra. For a much more detailed explanation 
of these results the reader is referred to \cite{RS}.
\smallskip

There is a version of the state-field  correspondence
in boundary conformal field theory that assigns a boundary 
field to each state in $\cH_{IJ}$. Hence we can read off 
{}from (\ref{bpart}) that the boundary primary field $\psi_j$ 
appears with multiplicity $ N_{Ij}^{\ \ J}$ in the boundary theory. 
The operator product expansion for two such primary fields is 
given by 
\beq
\psi^{LM}_i(x_1) \ \psi^{MN}_j(x_2) \ = \ 
\sum_{k} \ (x_1- x_2)^{h_i+h_j-h_k}\  
  \psi^{LN}_k (x_2) \ \Fus{M}{k}{i}{j}{L}{N}\ + \dots \ \ 
  \mbox{ for } \ x_1 < x_2 ,
\label{bOPE}
\eeq
where $F$ stands for the fusing matrix of the chiral algebra 
$\cA$. It is defined as a linear transformation that relates 
two different orthonormal bases in the space of conformal 
blocks (see \cite{MooSei} and Subsection 4.2 below) and it 
can be visualized through our Figure 1. 

\vspace{1cm}
\fig{Graphical description of the fusing matrix.}
{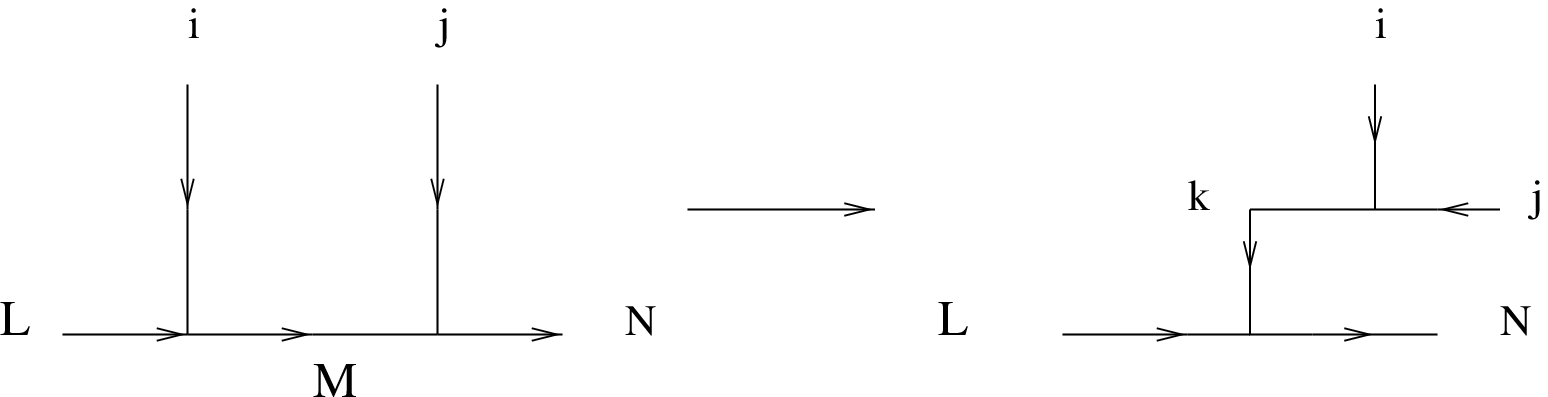}{15truecm}
\figlabel{\basic}
\vspace{1cm}

The formula (\ref{bOPE}) was originally found for minimal 
models by Runkel \cite{Run} and extended to more general 
cases in \cite{FFFS1, FFFS2, zuber}. Geometrically, 
boundary operator products describes the scattering of two open 
strings which are stretched between the branes $L,M$ and $M,N$, 
respectively, into an open string that stretches between 
$L$ and $N$. 

Note that for the relation between the coefficients of the boundary 
OPE and the fusing matrix it is crucial that boundary conditions 
and boundary fields are labeled with elements from the same set. 
This is no longer true for models with a `non-diagonal' (in the 
sense specified above) bulk modular invariant partition function.  
We shall see below how this can affect the boundary operator 
product expansions. The first examples of boundary OPEs for 
non-diagonal modular invariants were studied by Runkel in 
\cite{Run2}.  

\subsection{Boundary conditions for the orbifold} 
Suppose now that we want to discuss D-branes on an orbifold
of the original conformal field theory. Geometrically, one 
would like to understand these branes on the orbifold space 
through D-branes on the covering space. In such an approach,
a brane on the orbifold gets represented by several pre-images
on the covering space which are mapped onto each other by the 
action of the orbifold group. As we discussed in \cite{BruSch},
there is a large class of cases in which these geometric ideas
carry over to the construction of branes in exactly solvable
conformal field theories. 
\medskip

Our main assumption is that the orbifold action is induced by 
simple currents of the conformal field theory. Before we make 
this more precise, let us introduce some notations. Primaries
(or the associated conformal families) of a conformal field 
theory form a set $\cJ$. Within this set $\cJ$ there can be 
non-trivial elements $g \in \cJ$ such that the fusion product 
of $g$ with any other $j \in \cJ$ gives again a single primary 
$g \times j = gj \in \cJ$. Such elements $g$ are called {\em 
simple currents} and the set ${\cal C}$ of all these simple 
currents forms an abelian subgroup ${\cal C} \subset \cJ$. The 
product in ${\cal C}$ is inherited from the fusion product of 
representations. From now on, let us fix some subgroup $\Gamma 
\subset {\cal C}$. 
\smallskip

Through the fusion of representations, the index set $\cJ$
comes equipped with an action $\Gamma \times \cJ \rightarrow
\cJ$ of the group $\Gamma$ on labels $j \in \cJ$. Under this
action, $\cJ$ may be decomposed into orbits. The space of these
orbits will be denoted by $\cJ / \Gamma$ and we use the symbol
$[j]$ to denote the orbit represented by $j \in \cJ$. These 
orbits may have {\em fixed points}, i.e.\ there can be labels 
$j \in \cJ$ for which the following stabilizer subgroup $\S_j 
\subset \Gamma$
\beq\label{stab}
\S_{j} \ = \ \{ \ g \in \Gamma \ \mid\  g \cdot j \ = \ j\ \}
\eeq
is nontrivial. Up to isomorphism, the stabilizer subgroups
depend only on the orbits $[j]$ not on the choice of a 
particular representative $j \in [j]$, i.e.\ $S_j = S_[J]$.  
\smallskip

The last object we have to introduce is the so-called {\em 
monodromy charge} $Q_g(j)$ of a primary $j$ with respect to 
the simple current $g$. To this end we consider the 
following special matrix elements 
$$\Omega \vert{i}{k}{j} \ = \ \Br{i}{k}{i}{j}{0}{k} $$  
of the braiding matrix $B = B^{(+)}$ (see \cite{MooSei} for 
details). From these elements of the braiding matrix 
we inherit a map $\tQ_g(j)$ defined by 
\beq \label{Qdef} 
(-1)^{\tQ_g(j)} \ := \Omega \vert{j}{gj}{g} \ \ .
\eeq
Note that this specifies $\tQ_g(j)$ up to an even integer, 
i.e. $\tQ_g(j) \in \QR/ 2\QZ$. The monodromy charge 
$Q_g(j) := \tQ_g(j)\ \mod\ 1$ is only defined up to 
integers. Note that the latter is given by the standard
formula 
$$ Q_g(j) \ = \ h_j + h_{g} - h_{gj} \ \ \mod \ \ 1 \ \ . $$
Let us remark that $Q$ is conserved under fusion while 
this may not be the true for $\tQ$. In case the simple 
currents have integer conformal weight, the monodromy charge 
$Q_g(j)$ depends only on the equivalence class $[j]$ of $j 
\in J$. An orbit $[j]$ is said to be invariant, if $Q_g([j]) = 
Q_g(j) = 0$ for all $g\in \Gamma$. 
\smallskip

After this preparation, we can give a precise formulation of our main 
assumption on the partition function $Z^{\orb}(q)$ of the bulk theory that
we want to study. We assume that there exists some bulk theory to which 
Cardy's theory applies, an orbifold group $\Gamma$ within the group 
of all simple currents such that $Z^{\orb}$ is of the form 
\beq\label{intmod}
Z^{\orb}(q) \ = \ \sum_{j, Q_\Gamma ([j]) = 0 } \ |\, \S_{[j]}\, | \
   | \, \sum_{g\in \Gamma/\S_{[j]}}\ \chi_{gj} \, |^2  \ \ . 
\eeq
Note that this partition function does not have the simple form (\ref{Bpart}) 
so that Cardy's theory for the classification and construction of D-branes 
does not apply directly. 
\medskip   

As we discussed in \cite{BruSch}, an orbifold theory with bulk 
partition function of the form (\ref{intmod}) possesses 
consistent boundary theories which are assigned to orbits 
$[I]$ of labels $I$ that parametrize the boundary theories 
of the parent CFT. The open string spectra associated with 
a pair of such brane on the orbifold are given by 
\beq \label{Zproj}  
Z^{\orb}_{[I][J]}(q) \ = \  
\sum_{g,k} \ N_{I \,\, k}^{\ \ gJ} \chi_{k}(q)\ \ .
\eeq
This agrees precisely with the prediction from the geometric 
picture of branes on orbifolds. In fact, the $I,J$ can be 
considered as geometric labels specifying the position of the
brane on the covering space. To compute spectrum of two branes 
$[I]$ and $[J]$ of the orbifold theory, we lift $[I]$ to one of 
its preimages $I$ on the covering space and  include all the 
open strings that stretch between this fixed brane $I$ on the 
cover and an arbitrary preimage $gJ$ of the second brane $[J]$. 
\smallskip

We should remark that in many cases the boundary conditions 
$[I]$ can be further resolved, i.e.\ there exists a larger set 
of boundary theories such that $[I]$ can be written as a 
linear combination of boundary theories with integer 
coefficients. This happens whenever the stabilizer subgroup
$S_{[I]}$ is non-trivial (see \cite{FucSchI, FucSchII,FSW,BruSch} 
for more  details). Geometrically this corresponds to the fact 
that the CP-factors of branes at orbifold fixed points can carry 
different representations of the stabilizer subgroup.

\subsection{Boundary OPE for the orbifold} 
  
Restricting to unresolved D-branes, it is relatively easy to
give explicit expressions for the operator products of boundary
fields. Before we spell them out, let us have another look at 
eq. (\ref{Zproj}) and observe that for fixed $I,J,k$ there can 
be several group elements $g \in \Gamma$ such that $N_{I\, \, k}^{gJ} 
\neq 0$. We label these elements by a subscript $\epsilon$. While 
the range for $\epsilon$ depends only on $k$ and the orbits $[I],[J]$, 
the definition of the group elements $g_\e = g_\e \vert{k}{I}{J}\in 
\Gamma$ requires to fix representatives $I \in [I]$ and $J \in [J]$. 
If we shift these representatives along their orbits, the group 
elements behave according to   
$$ g_\e \vert{k}{I}{gJ} \ = \ g^{-1}\, g_\e \vert{k}{I}{J} \ \ 
   \mbox{ and }  \ \ 
   g_\e \vert{k}{gI}{J} \ = \ g  \, g_\e \vert{k}{I}{J} \ \ . 
$$ 
The group elements $g_\e$ appear in the following formula for the 
boundary operator product expansions in our orbifold theory,  
\beq \label{oope} 
 \Psi^{[L][M]}_{i,\e_1} (x_1) \ \Psi^{[M][N]}_{j,\e_2} (x_2) \ = \ 
   \sum_k \ (x_1-x_2)^{h_i+h_j-h_k} \Psi^{[L][N]}_{k,\e_{12}} (x_2)
   \ \Fus{g_1 M}{k}{i}{j}{L}{g_{12} N} \ + \ \dots  \  
\eeq
for $x_1 < x_2$. 
Here, $L,M,N$ are representatives of the orbits $[L],[M],[N]$ and 
the group elements $g_1,g_{12}$ in the fusing matrix $F$ are given 
by 
$$ 
 g_1 \ = \ g_{\e_1} \vert{i}{L}{M} \ \ \ , \ \ \  
 g_{12} \ = \ g_{\e_{12}}\vert{k}{L}{N} \ = \ g_{\e_1} 
 \vert{i}{L}{M} g_{\e_2} \vert
 {j}{M}{N} \ \ . 
$$  
Obviously, the expansions (\ref{oope}) in the orbifold theory are 
inherited from the operator products (\ref{bOPE}) of the theory 
on the covering space. In geometric terms we have singled out one
of the preimages $L$ of the branes $[L]$ and then described the 
scattering of open strings between strings on the orbifold through
strings stretching between various preimages of the other two 
branes $[M],[N]$. This prescription is independent of the choices 
we have made, provided that the charge $\tQ$ is conserved in the
sense  
\begin{equation} \label{assum2} 
 \tQ_g(i) + \tQ_g(j) - \tQ_g(k) \ = \ 0 \ \mod \ 2 \ \ \mbox{ for all } \ \ 
 i,j,k \ \mbox{ with } \ N_{ij}^k \neq 0 \ \ .
\end{equation}
More precisely, one can show that under the assumption (\ref{assum2})
our operator product expansions (\ref{oope}) obey the usual factorization
(or {\em sewing}) constrains \cite{Lew,PrSaSt,Run}. In the derivation 
one uses the following invariance of the fusing matrix   
$$ \Fus{g g_1 M}{k}{i}{j}{gL}{g g_{12}N} \ = \ 
      \Fus{g_1}{k}{i}{j}{L}{g_{12}N}
           (-1)^{\tQ_g(i) + \tQ_g(j) - \tQ_g(k)} \ \ . 
$$     
It is possible to relax the assumptions and to generalize 
the formula for the operator product expansions but for our 
purposes, eq.\ (\ref{oope}) will turn out to suffice.

\section{The  $\CN=2$ superconformal minimal models} 
\def\ov{\overline} 
\def\sfp{{\sf p}} 
\def\MM{{\rm MM}} 
\def\id{{\rm id}} 

The $\CN=2$ minimal models are the basic building blocks
for Gepner models. More precisely, the latter are obtained 
{}from a product of $\CN=2$ minimal models by orbifold techniques.
Following the general strategy of the previous section it is
therefore essential to understand the open string sector of 
$\CN=2$ minimal models, including the OPE of the open string 
operators. As we have seen, the operator products are determined
by the fusing matrix. It is the main aim of this section to 
compute fusing matrix of $\CN=2$ minimal models. This will be
achieved through the coset construction.

\subsection{The coset construction for $\CN=2$ minimal models}

The $\CN=2$ minimal models $\MM_k$ have a coset realization of the 
following form 
\beq \label{coset}
\MM_k \ = \ \frac{\su \times \U_4}{\U_{2k+4}}
\eeq
where $\su$ is a level $k$ affine current algebra and $\U_{2N}$ 
stands for an extension of the $U(1)$ current algebra that is 
generated by the exponentials $W^{(2N)}_\pm \ = \ \exp(\pm i {\sqrt{2N}} 
X(z))$ along with the current $J^{(2N)}(z) = J(z)$.  We will denote the 
generators of the $\su$ current algebra by $E(z), F(z)$ and $H(z)$. As 
usual, $H(z)$ is associated with the Cartan subalgebra of $su(2)$ while 
$E(z)$ and $F(z)$ come with the raising and lowering operators, respectively. 
The space of ground states of the $\su$ representations is spanned by 
vectors $|l,n\rangle, |n| \leq l,$ which are eigenstates of the zero
mode $H_0$ with eigenvalue $n$. The coset construction of $g/g'$ 
requires to  embed the denominator $g'$ into the nominator $g$. In 
the case at hand this is achieved through the identification 
$$  J^{(2k+4)} (z) \ = \  H(z) + J^{(4)}(z) \ \ . $$ 
The Virasoro field of the coset theory is then given by the difference
$T^{g/g'}(z) = T^g(z) - T^{g'}(z)$ which has central charge $c^{g/g'} 
= c^g - c^{g'}$. These formulas specialize to 
\beqa  T^{\MM_k}(z) & = & T^{\su}(z) + T^{\U_4}(z) - T^{\U_{2k+4}}(z) 
     \\[2mm] \mbox{ with } & & c^{\MM_k} = c^{\su} \ +1 \ -1 \  = 
       \frac{3k}{k+2}\ \ .
\eeqa
Representations of the coset algebra can be realized on the representation
spaces $\cH_\lambda$ of the theory $g$ in the nominator. The embedding of 
$g'$ into $g$ defines an action of the denominator on $\cH_\lambda$. With 
respect to this action $\cH_\lambda$ decomposes according to  
\beq \label{dec}
{\cal H}_\lambda \ = \ \bigoplus_{\lambda'} {\cal H}_{\lambda \lambda'}
\otimes {\cal H}_{\lambda'}
\eeq
where $\cH_{\lambda'}$ are sectors of the denominator $g'$. The spaces
$\cH_{\lambda \lambda'}$ which appear in this decomposition are acted 
upon by all fields which commute with the fields of the denominator
theory and hence they carry representations of the coset algebra $g/g'$. 
In general, it is difficult to decide, which representations $\lambda'$ 
of $g'$ occur for given $\lambda$ an which pairs of $\lambda,\lambda'$ 
give rise to the same representation for $g/g'$. To determine such 
selection rules and identifications is rather difficult, in general, 
but simple currents can help with this task. In this approach one 
constructs the largest group $\Gamma_\id$ of integer weight simple 
currents from the theory $g \oplus g'$ such that all the pairs 
$\lambda, \lambda'$ that appear in the sum (\ref{dec}) have vanishing 
monodromy charge $Q_g (\lambda) + Q_g(\lambda') = 0 \ \mod \ 1$. 
Elements of $\Lambda_\id$ are called {\em identification currents} 
since their action on pairs $\lambda,\lambda'$ generates orbits of 
identical representation spaces $\cH_{\lambda\lambda'}$ for the 
coset theory. Additional complications arise, if not all of these 
orbits have equal length. The fixed point resolutions required in 
such cases have been analyzed in \cite{multifsI,multifsII}. We will
not discuss this process here, since there are no such fixed points
for $\CN=2$ minimal models. 

The representations of the nominator in the cosets (\ref{coset}) are 
labeled by pairs $(l,s)$ with $l = 0,1,\dots,k$ and $s = -1,0,1,2$. 
Under fusion, the labels $l$ obey the usual $\su$ fusion rules while 
the four choices for $s$ get identified with elements of $\QZ_4$. 
Likewise, we enumerate the sectors of the denominator $\U_{2k+4}$ 
by $m = -(k+1), -k, \dots, k+1,k+2$. They form the abelian group
$\QZ_{2k+4}$. Among the triples $(l,m,s)$ labeling representations
of $\su \oplus \U_4 \oplus \U_{2k+4}^*$ there is the generator
$\gamma = (l=k,m=k+2,s=2)$ of the identification group $\Gamma_\id 
\cong \QZ_2$. It maps the element $(l,m,s)$ to $\gamma (l,m,s) = 
(k-l,m+k+2,s+2)$. The following formula for the monodromy charge  
\beq
Q_\gamma (l,m,s) \ = \ h_{(k,2,k+2)} + h_{(l,s,m)}- h_{(k-l, m+k+2, s+2)}
\ = \ \frac{l+m-s}{2}\ \mod \ 1 \ 
\eeq
encodes the selection rules through the the requirement $Q_\gamma = 0$ 
and hence it leads us to the decomposition 
\beq \label{dec2}
    \cH^\su_l \otimes \cH_s^{\U_4}   \ = \ \bigoplus_{m, l+m+s \ 
               \mbox{\rm {\small even}}} 
   \cH^{\MM_k}_{(l,m,s)} \o \cH^{\U_{2k+4}}_m \ \  .
\eeq 
In addition, we recover the well-known field identification of 
$\CN=2$ minimal models from the action of the identification current 
$\gamma$. It states that $\cH_{(l,m,s)}$ and $\cH_{(k-l,m+k+2,s+2)}$ 
carry the same representation of the coset $\MM_k$. 
\medskip

So far, our remarks on the coset construction have been fairly 
standard. But our construction of the fusing matrix requires additional
input. Namely, we have to understand in detail, how the ground states
$|l,m,s\rangle \otimes |m\rangle$ of $\cH_{(l,m,s)} \otimes \cH_m$ 
are realized within the representation spaces $\cH_l  \otimes \cH_s$ 
of the nominator theory. We know from (\ref{dec2}) that each sector
$(l,s)$ of the nominator theory must contain $k+2$ such ground states.   
To find these ground states we fix $l,s$ and choose $m$ such 
that $l+m+2$ is even. Within the subspace 
\beq \label{Hlsm} \cH_{ls}^{(m)} \ = \ (\cH_l \o \cH_s)^{(m)} \ := \ 
    \{ \psi \in \cH_l \o \cH_s \ |\ 
   e^{\frac{2 \pi i}{\sqrt{2k+4}} J_0^{(2k+4)}} \, \psi \ = \ 
   e^{\frac{2\pi i m}{2k+4}} \, \psi \ \} 
\eeq 
we search then for eigenstates $\psi^{(m)}_{ls}$ of $L^\su_0 + L^{\U_4}_0$ 
with minimal eigenvalue.    

Some of these states are easily identified. These are the states 
which are realized in terms of ground states of the nominator
theory,  
\beq
\psi^{(m)}_{l\, s} \ = \ |l,m,s\rangle \otimes |m\rangle \ = \ |l, n=m-s\rangle \otimes 
                                      |s\rangle\ \ 
\eeq
where $m$ is restricted by $|m-s|\leq l$. In this way we have realized 
all fields from the so-called standard range of $\CN=2$ minimal models, 
\beq \label{srange} 
 l \ \leq\  k\ , \ \ \quad |m-s| \ \leq \ l, \ , \ l + m + s 
 \ \mbox{\rm even} \ \ . 
\eeq
For these fields, the following formula for their conformal weights 
holds exactly (not just up to an integer),  
$$ h_{(l,m,s)} \ = \ \frac{l(l+1) - m^2}{4(k+2)} + \frac{s^2}{8}
   \ \ . $$    
But the $l+1$ states we have found do not exhaust the ground states 
of the denominator theory. Additional states can be constructed with
the help of 
$$
E^\nu_{-1} |l,l\rangle \ \ , \ \ F^\nu_{-1} |l,-l\rangle  \ \in 
\cH_l  \ \ \mbox{ for } \ \ \nu = 1,2,\dots \ \ . 
$$
These states carry the charge $n = l + 2\nu$ or $n = -l-2\nu$, 
respectively. When combined with appropriate states from $\cH_s$ 
(not necessarily the ground state) they furnish all the ground 
states for the denominator theory of the coset (\ref{coset}). 
Spelling out the complete list is somewhat involved and since 
we do not need the explicit formulas in the following, we shall 
content ourselves with these sketchy remarks.
\bigskip

Before we conclude this subsection, let us briefly discuss some
important simple currents of the $\CN=2$ minimal model. The 
fusion rules for $(l,m,s)$ which we have described above imply 
that e.g.\ $(0,1,1)$ and $(0,0,2)$ are both simple currents. 
They are of special interest in the context of Gepner models 
and will be used to generate our simple current group $\Gamma$. 
$(0,1,1)$ is the spectral flow by $1/2$ unit and $(0,0,2)$ the 
world-sheet supersymmetry generator. $(0,0,2)$ is a simple 
current of order $2$ and can be used to combine the world-sheet 
fields into supermultiplets. The order of the simple current 
$(0,1,1)$ is model dependent. To see this, we apply the current 
$2k+4$ times to the identity. This will lead us back to the 
identity whenever the level $k$ is even. Since $(0,0,2)$ is 
nowhere on this orbit, $(0,0,2)$ and $(0,1,1)$ together generate 
the simple current group $\Gamma = \Gamma_k = \QZ_{2k+4} \times 
\QZ_2$ for even level $k$. When $k$ is odd, however, we reach 
the field $(0,0,2)$ after $2k+4$ applications of $(0,1,1)$ and 
hence we have to apply the simple current $(0,1,1)$ another 
$2k+4$ times. In this case, the orbit contains the label
$(0,0,2)$ and hence our orbifold group is $\Gamma = \Gamma_k = 
\QZ_{4k+8}$ for odd $k$. 

The orbits for the action of $\Gamma_k$ on the set $\cJ$ depend 
once more on the parity of $k$. If $k$ is odd, the orbifold group 
$\Gamma_k$ acts freely so that all orbits have length $4k+8$. For 
even level $k$, however, we generate short orbits of length $2k+4$ 
whenever we start from a field $(l,m,s)$ with $l = k/2$ because 
the label $l=k/2$ is invariant under field identification. The 
stabilizer for these short orbits is a subgroup $\QZ_2 \subset 
\Gamma_k$.

\subsection{The fusing matrix}

As discussed in the previous section, the coset construction we will
be using below involves two basic building blocks, one of them being 
the ${\widehat{SU}}(2)_k$ Kac-Moody algebra while the other is simply 
some abelian $U_{2N}$-theory. Here we shall briefly describe the fusing 
matrices of these two theories before we turn to the fusing matrix of 
$\CN=2$ superconformal minimal models.
\medskip

Let us start with the discussion of $\widehat{SU}(2)_k$ to illustrate 
the basic steps that go into the construction of fusing matrices. The 
reader may profit from the more comprehensive treatment 
in \cite{MooSei}.   

To any three given labels $l,l_s,l_t$ such that $l_t$ is contained 
in the fusion product of $l$ and $l_s$ there is assigned an intertwiner 
$$\phi \vvert{\ l}{l_t}{l_s}(z): \ \cH_{l} \o \cH_{l_s} \ \rightarrow 
\ \cH_{l_t}$$ 
that intertwines  the natural action of the affine Kac-Moody algebra
on the involved spaces. For $\cH_l\o \cH_{l_s}$ this action is through 
the so-called fusion product and it depends on the co-ordinate $z$. 
Let us pick an orthonormal basis $\{ |l,\nu\rangle; \nu \in -2k+l + 2
\QN^0\}$ of vectors in $\cH_{l}$ such that $|l\rangle = |l,-l\rangle $ 
is primary. It will also be convenient below to reserve $|l,n\rangle, 
|n|<2k-l$ for the vectors 
$$ |l,\nu\rangle \ = \ \left\{ \begin{array}{rl} 
   F^{-\frac12(l+\nu)}_{-1}\, |l,-l\rangle  & \mbox{ for } \ \nu = -2k+l\, \dots, -l-2\\[2mm]
   |l,\nu\rangle & \mbox{ for } \ \nu = - l, \dots, l   \\[2mm]  
   E^{\frac12(\nu+l)}_{-1}\ |l,\ l\rangle  & \mbox{ for } \ \nu = l+2, 
   \dots, 2k+l\ \ \ \ \ .  
\end{array} \right.  
$$ 
To each state $|l,\nu\rangle$ from the basis we assign a vertex operators by 
$$\phi  \vvert{l,\nu}{l_t}{l_s}(z) \ := \ \phi  \vvert{l,\nu}{l_t}{l_s}
[|l,\nu\rangle; . ](z): \ \cH_{l_s}\  \rightarrow \ \cH_{l_t}\ \ . $$ 
These vertex operators are uniquely determined by their intertwining 
properties, up to a common normalization that we fix by requiring 
that 
\beq \langle l_t|\ \phi \vvert{l,l_s-l_t}{l_t}{l_s}(1)\ |l_s\rangle 
\ = \ 1\ \ . \label{vertnorm}
\eeq 
Now we can finally define the fusing matrix through an operator 
product expansion of chiral vertex operators,  
$$ \phi \vvert{l_3,\mu}{l}{l_{12}}(z_1)\  \phi \vvert{l_2,\nu}{l_{12}}
{l_1}(z_2) \ = \ \sum_{l_{23},\rho}\ \Fus{l_{12}}{l_{23}}{l_2}{l_3}
{l_1}{l} \ \phi \vvert{l_{23},\rho}{l}{l_1}(z_2) \ \langle l_{23}, \rho| 
\phi \vvert{l_2,\nu}{l_{23}}{l_3}(z_{12})|l_3,\mu\rangle \ \ , $$ 
where $z_{12} = z_1-z_2$. Explicit formulas for the fusing matrix 
can be found in the literature. One can also compute the 
3-point functions on the right hand side. When all the involved 
states $|l_2,\nu\rangle, |l_3,\mu\rangle$ and $|l_{23},\rho\rangle$ 
are taken from the lowest energy subspaces(i.e.\ $|\nu| \leq l_2, |\mu| 
\leq l_3$ and $|\rho| \leq l_{23}$), these amplitudes can be 
expressed in terms of  Clebsch-Gordan coefficients for the finite 
dimensional Lie algebra $su(2)$,      
\beq \langle l_{23},n_{23}| \ 
\phi \vvert{l_2,n_2}{l_{23}}{l_3}\ |l_3,n_3\rangle
\ = \ \CG{l_3}{l_2}{l_{23}}{n_3}{n_2}{n_{23}} \ \ \ \ 
    \mbox{ with }\ \ \ \ \CG {\ l_3}{\ \ l_2}{\ l_{23}}
{-l_3}{l_3- l_{23}}{-l_{23}} \ = \ 1  \ \ .  
\label{transCG} \eeq
The normalization condition for the Clebsch-Gordan coefficients
is required to match with the corresponding condition (\ref{vertnorm}) 
for  chiral vertex operators. Relation (\ref{transCG}) implies that 
the coefficient of the leading singularity in the operator product 
expansion of two primary fields is given by the fusing matrix.   
\medskip 

The algebras $U_{2N}$ are much simpler to describe. They have
central charge $c=1$ and sectors labeled by $m = -N+1, \dots, N$
with primaries of conformal dimension $h_m = m^2/4N$. The 
fusion rules are just given by the composition in $\QZ_{2N}$.
We will usually assume that our labels $m$  run through the 
allowed range $-N+1 \leq m \leq N$. When two such integers 
$m_1$ and $m_2$ are added, their sum does not necessarily lie
in the same range. To cure this problem it will be useful to 
work with another sum $\hp$ which is defined such that 
$m_1 \hp m_2$ is the unique integer between $-N+1$ and $N$ 
such that $m_1 \hp m_2 = m_1 + m_2$ mod $2N$. 
\smallskip

Again, we are mainly interested in the fusing matrix of $U_{2N}$. 
This time we can compute it in detail using the general recipes 
we have sketched before. The basic fields in the chiral algebra 
of the $U_{2N}$-theory may be obtained from a single free bosonic 
field through the expression 
$$ W_\nu(z) \ :=\ \ :\!e^{i \nu \sqrt{2N} X(z)}\!: \ \ \ \mbox{ where } 
\ \ \ \nu \in \QZ$$
Using the familiar OPE of such normal ordered exponentials it is 
easy to see that these fields are local with respect to each other, 
i.e.\ that their correlation functions are meromorphic throughout 
the complex plane. 

Similarly, we would like to realize the vertex operators in terms
of normal ordered exponentials. This is possible by means of the 
formula
\beq \phi\vvert{m,\mu}{m_t}{m_s}(z) \ := \ \
   :\!e^{i \frac{m+\mu 2N} {\sqrt{2N}} X(z)}\!: 
    \ (-1)^{\frac{m (\sfp - m_s)}{2N}}|_{\cH_{m_s}}  
   \ \ , 
\label{Uvert} \eeq
where $\sfp= J_0 \sqrt{2N}$ is obtained from the zero mode of 
the current $J(z) = i \partial X(z) $ by a simple rescaling. The 
vertex operator is restricted to the subspace $\cH_{m_s}$ in 
the full state space of the free bosonic field which carries 
the $m_s$-representation of the $U_{2N}$ algebra. Formally, it 
can be defined on an eigenspace of the operator $\exp(2\pi i 
\sfp/ 2N)$ with the eigenvalue $\exp(2\pi i m_s/2N)$. Note that 
by construction the operators defined in eq. (\ref{Uvert}) 
provide maps between different sectors of the $U_{2N}$-theory. 
The $\sfp$-dependent factor was introduced to guarantee that 
the operators (\ref{Uvert}) have trivial braiding with respect 
to all the chiral fields $W_\nu(z)$, thereby turning them into 
true vertex operators. Setting $\mu = 0$ in eq.\ (\ref{Uvert}),  
we obtain primaries with respect to the extended chiral algebra. 
All other vertex operators in eq. (\ref{Uvert}) are $U_{2N}$-%
descendants of primary fields. Finally, we have subtracted the 
integer $m_s$ from $\sfp$ to ensure that the transition 
amplitudes satisfy the normalization condition 
$$
 \langle m_t |\  \phi \vvert{m,\mu}{m_t} {m_s}(1)\ 
 |m_s\rangle \ = \ \delta_{m_t,m+\mu 2N + m_s}\ \ . 
$$ 
This is analogous to the normalization condition we have 
imposed on the 3-point functions in the $SU(2)$-theory above. 
\smallskip

The computation of the operator product expansion of any two 
such vertex operators (\ref{Uvert}) is rather straightforward 
and it gives the following fusing matrix,  
\beq \Fus{m_{12}}{m_{23}}{m_2}{m_3}{m_1}{m} 
   \ = \ 
   (-1)^{\frac{m_3}{2N}(m_1 + m_2 - m_{12})} \ 
   \delta_{m_{12}, m_1 \hp m_2}\  
   \delta_{m_{23}, m_2 \hp m_3} \ 
   \delta_{m, m_{12} \hp m_3}\ \ . 
\label{UFus} \eeq
One can also use the vertex operators (\ref{Uvert}) to compute 
the braiding matrix matrix of $U_{2N}$-theory and then check 
that these objects satisfy all the polynomial equations of 
\cite{MooSei}. Our solution of these equations actually differs 
{}from the one described in Appendix E of \cite{MooSei} by a 
`gauge transformation' with $\lambda(i,j) = (-1)^{ij}$. 
\bigskip

Let us now turn towards our ultimate goal to determine the 
fusing matrix of the $\CN=2$ superconformal algebra. For the latter, 
we will be using its $\widehat{SU}(2)_k \times U_4/ U_{2k+4}$-coset 
presentation discussed before.

The construction requires to find the vertex operators of 
the $\CN=2$ minimal models. Their products with the vertex operators
of the $\U_{2k+4}$ theory in the denominator can be realized on the 
spaces $\cH_{ls}^{(m)}$ introduced in eq.\ (\ref{Hlsm}),   
\beq \label{MMvert}
\phi \vvert{(l,s,m)}{(l_t,s_t,m_t)}{(l_s,s_s,m_s)} 
   \ \phi \vvert{m}{m_t}{m_s}\, :\, \cH_{l_s\, s_s}^{(m_s)}  \ \to\  
    \cH_{l_t\, s_t}^{(m_t)} \ \ . 
\eeq
Using the explicit embedding of ground states of $\MM_k \oplus \U_{2k+4}$ 
into the representation space $\cH_l \o \cH_s$ that we sketched in the 
previous subsection, we decompose the operators (\ref{MMvert}) into 
products of vertex operators for the nominator theory,  
\beqa \label{vdec}  
\phi \vvert{(l,s,m)}{(l_t, s_t, m_t)}{(l_s, s_s, m_s)} 
\phi \vvert{m}{m_t}{m_s}  & = &   \inorm{l_s}{l\ }{l_t}{m_s}{m\ }{m_t}{s_s}{s\ }{s_t} 
   \phi \vvert{\psi^{(m)}_{ls;1}}{l_t}{l_s} 
   \phi\vvert{\psi^{(m)}_{ls;2}}{s_t}{s_s}\\[4mm]
\mbox{where } \ \ \ \ \   \norm{l_s}{l\ }{l_t}{m_s}{m\ }{m_t}{s_s}{s\ }{s_t}
    & := & \vev{\psi^{(m_t)}_{l_t s_t}| \phi \vvert{\psi^{(m)}_{ls;1}}
     {l_t}{l_s} 
    \phi \vvert{\psi^{(m)}_{ls;2}}{s_t}{s_s} |\psi^{(m_s)}_{l_s s_s}}\ \ . \nn
\eeqa
Here, $\psi^{(m)}_{l\, s} \in \cH_l \o \cH_s$ denotes the vector introduced 
in the text after eq.\ (\ref{Hlsm}). This vector can be split into a product 
$\psi^{(m)}_{ls;1} \in \cH_l$ and $\psi^{(m)}_{ls;2} \in \cH_s$ of components
in the representations spaces of the $\su$ and the $\U_4$ algebras, 
respectively. If all the three triples $(l,m,s), (l_s,m_s,s_s), (l_t,m_t,s_t)$ 
are in the standard range (\ref{srange}), the normalization in eq.\ 
(\ref{vdec})
can be spelled out more explicitly  with the help of eq.\ 
(\ref{transCG}),         
\beq \label{norm} 
    \norm{l_s}{l\ }{l_t}{m_s}{m\ }{m_t}{s_s}{s\ }{s_t} \ = \
   \CG{l_s}{l}{l_t}{n_s}{n}{n_t} \ \ .  
  \eeq
Here $n_\alpha = m_\alpha - s_\alpha$. We can now employ the formula 
(\ref{vdec}) to compute the fusing matrix of the superconformal algebra
in terms of the fusing matrices of the building blocks of the coset 
construction (\ref{coset}). It is given by 
\beqa \label{mmfus}
\Fus{(l_{12},m_{12},s_{12})}{(l_{23},m_{23},s_{23})}
{(l_2,m_2,s_2)}{(l_3,m_3,s_3)}{(l_1,m_1,s_1)}{(l,m,s)} 
 & = & (-1)^{\frac{s_3}{4}(s_1 + s_2 - s_{12})}\ \delta_{s_{12},s_1\hp s_2} 
  \  \delta_{s_{23},s_2 \hp s_3}\ \delta_{s , s_{12} \hp s_3}
  \nn \\[2mm] & & \hspace*{-7.9cm}  
   (-1)^{-\frac{m_3}{2k+4} (m_1 + m_2 - m_{12})} \ \delta_{m_{12},m_1\hp m_2} 
  \ \delta_{m_{23},m_2 \hp m_3}\ \delta_{m,m_{12} \hp m_3}\ 
  \Fus{l_{12}}{l_{23}}{l_2}{l_3\ }{l_1}{l} 
  \ \frac{\norm{l_3}{l_2}{l_{23}}{m_3}{m_2}{m_{23}}{s_3}{s_2}{s_{23}} \norm{l_1}
   {l_{23}}{l}{m_1}{m_{23}}{m}{s_1}{s_{23}}{s}} {\norm{l_1}{l_2}{l_{12}}{m_1}
   {m_2}{m_{12}}{s_1}{s_2}{s_{12}} \norm{l_{12}}{l_3}{l}{m_{12}}{m_3}{m}
    {s_{12}}{s_3}{s}}\, .
\nn 
\eeqa
If all the triples are in the standard range, then one can use eq.\ 
(\ref{norm}) to simplify this expression. However, in general it is 
not possible to bring all the involved triples into the standard
range in a way that is consistent with the fusion rules and the
factorization into the $SU(2)$ and $U(1)$ contributions. This is
because the triples $(l,m_1 \hp m_2, s_1 \hp s_2)$ with $l = 
|l_1 - l_2|, \dots, \mbox{\rm min}(l_1 + l_2, 2k-l_1-l_2)$ need
not be in the standard range even if $(l_1,m_1,s_1)$ and $(l_2, 
m_2,s_2)$ are.

\subsection{The OPE in a single $\CN=2$ minimal model}

In passing we would like to write down the boundary operator
product expansions for a single minimal model for the case
where Cardy's analysis applies, i.e. with the bulk 
modular invariant given by charge conjugation. As pointed out 
in Section 3, there exists a scaling of the boundary fields 
such that the OPE coefficients of the boundary OPE are equal 
to the fusing matrix. Hence, together with the results of the 
previous subsection we know all boundary structure constants. 
The expressions simplify, if we rescale all the boundary fields
to absorb some of the normalizations defined in eq.\ (\ref{vdec}). 
In formulas, the appropriate rescaling of the boundary states is 
given by
\beq\label{rescale}
\psi^{(LMS)(L'M'S')}_{(l,m,s)} \ \mapsto \psi^{(LMS)(L'M'S')}_{(l,m,s)}
\inorm{L'}{l}{L}{M'}{m}{M}{S'}{s}{S}
\eeq
For these rescaled vertex operators, the operator product expansion 
(\ref{bOPE}) then takes the form 
\beqa
&& \hspace*{-1cm} \psi^{(LMS)(L'M'S')}_{(lms)} (x_1) \ 
\psi^{(L'M'S')(L'' M'' S'')}_{(l'm's')} (x_2)
\ = \ 
\sum_{l'', m'', s''} \ (x_1-x_2)^{h + h'-h''}\ 
 \norm{l}{l'}{l''}{m}{m'}{m''}{s}{s'}{s''}\ \times \nn \\[2mm] 
&& \hspace{2cm} \Fus{l''}{L}{l}{l'}{L}{L''}\,  \Fus{s''}{S}{s}{s'}{S}{S''}
\ \left( \Fus{m''}{M}{m}{m'}{M}{M''}\right)^{-1}
\ \psi^{(LMS)(L''M''S'')}_{(l''m''s'')}(x_2) \ 
\label{MMope} 
\eeqa
whenever $x_1 < x_2$. This means that the structure constants of 
the operator product expansion equal a product of F-matrices 
times a normalization factor. The latter has no dependence on 
the boundary conditions. 

\section{Boundary OPE branes in Gepner models}

This section contains our main result, namely the boundary 
operator product expansions for arbitrary boundary fields
in a theory describing untwisted A-type branes in Gepner 
models. After a very brief introduction to Gepner models
we will recall the untwisted boundary states of \cite{RS}. 
The boundary operator product expansions are presented in 
the last subsection. 
 
\subsection{The Gepner model in the bulk}

The plan is to apply the general theory outlined
above to an important class of examples, namely to Gepner models
\cite{Gepn1,Gepn2} (see also \cite{Gree} for a review). These 
are exactly solvable CFTs which are used to study strings moving
on a Calabi-Yau manifold at small radius \cite{wittph}. Their 
construction employs an orbifold-like projection starting from 
a tensor products of $r$ $\CN=2$ minimal models. In our 
presentation we shall assume that there are $d=2$ complex, 
transverse, external dimensions with the appropriate ghost
sector. In order to get a consistent string background a GSO
projection has to be performed. This means that we project
on odd integer $U(1)$ charge in the full theory. In the internal
part we are projecting on integer charges. The GSO projected partition
function is of simple current type, where the simple current
is given by the spectral flow operator. To describe this more
explicitly we need some further notation.

Let us introduce the following vectors  
$$
{\mathbf \lambda} := (l_1,\ldots,l_r)\quad {\rm and}\quad 
{\mathbf \mu}:= (s_0; m_1,\ldots,m_r;s_1,\ldots,s_r)
$$ 
to label the representations $(l_j,\,m_j,\,s_j)$ of the individual 
minimal models and the representations $s_0=0,2,\pm1$ of a single 
$SO(2)$ current algebra at level $k=1$ that comes with the two 
complex fermions in the non-compact directions. The associated 
product of characters $\chi^{l_i}_{m_i,s_i}$ and $\chi_{s_0}$ is 
denoted by $\chi^{{\mathbf \lambda}}_{{\mathbf \mu}} (q)$.  
\smallskip

Next, we introduce the special $(2r+1)$-dimensional vectors $\beta_0$ 
with all entries equal to 1, and $\beta_j$, $j=1,\ldots,r$, having 
zeroes everywhere  except for the 1st and the $(r+1+j)$th entry which 
are equal to 2. These vectors stand for particular elements in the 
group $\BZ_4 \ti \prod_i \Gamma_{k_i}$. It is easily seen that they 
generate a subgroup $\Gamma = \BZ_K \ti \BZ_2^r$ where $K:= {\rm lcm}
(2k_j+4)$. Elements of this subgroup will be denoted by $\mathbfn 
\nu = (\nu,\nu_1, \dots, \nu_r)$. The monodromy charge of a pair 
$(\mathbf \lambda,\mathbf \mu)$ is 
\beqa  Q_{\smathbfn \nu} (\mathbf \lambda,\mathbf \mu) & = & \nu  
         \beta_0 \cdot {\mathbf \mu} + \sum_{i=1}^r \ \nu_i 
         \beta_i \cdot {\mathbf \mu}\ \  \mbox{\rm mod} \  1\  
     \label{Qtens}   \\[3mm] 
\mbox{where} \ \ \ \ \ \ 
\beta_0 \cdot {\mathbf \mu} &:=& 
- {s_0\over4} - \sum_{j=1}^r {s_j\over4} 
+ \sum_{j=1}^r {m_j\over 2k_j+4}\ , \\[2mm]
\beta_j \cdot {\mathbf \mu} &:=& -  {s_0\over2} 
 -{s_j\over2} \ .
\eeqa
The orbifold group $\Gamma$ acts on the labels $\mathbf \lambda$ 
and $\mathbf \mu$ in the obvious way. There appear orbits of 
maximal length $K 2^r$ and short orbits of length $K 2^{r-1}$. 
The latter are characterized by the property that $\mathbf 
\lambda = (l_1,\dots, l_r)$ satisfy $l_i = k_i/2$ for all 
$i$ such that $2k_i+4$ is not a factor in $K/2$. 
\medskip

When we consider the  full theory, the field generating the $\BZ_K$-%
symmetry contains a factor from the ghost sector and the space-time 
part of the spin field $S^\alpha$. In the $\pm1/2$ picture, the 
operator was spelled out in eq.\ (\ref{sf}) above.  It is  a simple 
current and its internal part agrees with the simple current $\beta_0$ 
in the tensor product considered above. 
Since the operator (\ref{sf}) has total weight one, it can be added to 
the chiral algebra and we can use the formula (\ref{intmod}) to determine
the partition function of the orbifold theory. The formula requires to 
determine invariant orbits, i.e.\ orbits with vanishing monodromy charge. 
Taking the OPE of the spectral flow (\ref{sf}) with a vertex operator that 
represent space-time scalars of the theory gives the monodromy charge
$$
\tilde Q_{\smathbfn \nu} (\mathbf \lambda,\mathbf \mu) \ = \ \nu  
         \left( \frac{\beta_0 \cdot {\mathbf \mu}}{2} + \frac12 
         \right) + \sum_{i=1}^r \ \nu_i 
         \beta_i \cdot {\mathbf \mu}\ \  \mbox{\rm mod} \  1\ \ . 
$$
$\tilde Q$ effectively replaces the monodromy charge $Q$ introduced 
in eq.\ (\ref{Qtens}). The orbits of vanishing monodromy charge are those 
of odd integer $U(1)$ charge. 
\smallskip

To write a partition function of physical states one has to 
extract the physical degrees of freedom. This can be done 
by a projection onto light-cone variables which removes, in 
particular, the ghost sector. In practice, the light-cone degrees
of freedom may be read off directly in the canonical ghost 
pictures.

An important reason to include ghosts is that the fields of the 
theory acquire the right commutation properties. One would like 
to incorporate this feature in the physical theory. Therefore, 
in the partition function, the fields are counted with a 
ghost-charge dependent phase factor $\exp (2\pi i q_{ghost})$. 
This means that the states with half-integer ghost charge, i.e.\ 
the RR-states,  contribute with a negative sign. 
\smallskip

The partition function in the light cone gauge is then given by
$$
Z^{(r)}_G (\tau,\bar \tau) = {1\over2} 
{({\rm Im}\,\tau)^{-{2}}\over |\eta(q)|^{2}}
\sum_{\mathbf \lambda,\mathbf \mu; \tilde Q (\mathbf \lambda,\mathbf \mu) = 0}
\ \sum_{\nu,\nu_j} \ \
(-1)^{\nu} \ \,\chi^{{\mathbf \lambda}}_{{\mathbf \mu}} (q)\ 
\,\chi^{{\mathbf \lambda}}_{{\mathbf \mu}+\nu\beta_0+\nu_1\beta_1+\ldots 
+\nu_r\beta_r} (\bar q)\ \ . 
$$
The sign is the usual one occurring in (space-time) fermion one-loop 
diagrams. The $\tau$-dependent factor in front of the sum accounts 
for the free bosons associated to the $2$ physical transversal 
dimensions of flat external space-time, while the $1/2$ is simply 
due to the field identification mentioned above. Except for these 
modifications,  the formula for $Z_G$ is the same as eq.\ 
(\ref{intmod}). Elements 
$g = \nu \beta_0 + \dots \nu_r \beta_r$ of the orbifold group 
$\Gamma$ are labeled by $\nu,\nu_i$ so that the second sum is 
over the full group $\Gamma$. Short orbits appear twice in the 
summation and give rise to an extra factor of $2$ which is the 
order of the corresponding stabilizer subgroup. Since our orbifold 
group $\Gamma$ is abelian, we used additive notation for the action 
of elements $g \in \Gamma$ on the labels $\mathbf \lambda, 
\mathbf \mu$.

\subsection{Boundary states in Gepner models}

Quite generally, we can construct a set of D-branes on
orbifolds by projecting down D-branes on the covering
space \cite{DM}. Let us explain how to apply these methods in the 
context of 
string compactification based on $\CN =2$ superconformal
field theories. Gepner models can then be discussed as
a  very special case
(the underlying $\CN =2$ CFT is rational) of such theories.
We will point out a few features which 
are common to branes in arbitrary $\CN=2$
theories, in particular they are not restricted to the Gepner point
of a given compactification.
To construct a theory with space-time supersymmetry, we
have to perform a GSO-projection, projecting out
all fields which are not local with respect to the supersymmetry
operator (\ref{sf}). 
We will mainly concentrate
on the internal part of (\ref{sf}), which consists of
the spectral flow operator in the internal dimensions.
\smallskip

To discuss D-branes which preserve some of the space-time
supersymmetry, we have to impose boundary conditions on the
spectral flow. They read  
\begin{eqnarray}
{\rm A-type:} \quad \partial X_L &=& - \partial X_R, \quad
 e^{iX_L} \ =  \ e^{-iX_R} \\ \nonumber
{\rm B-type:} \quad \partial X_L &=& \partial X_R, \quad
 e^{iX} \ = \ e^{iX_R} \ \ . 
\end{eqnarray}
This means that they are just Dirichlet/Neumann (for A-type/B-type)
conditions along the spectral flow. In the following, we discuss
A-type/Dirichlet conditions. Let us  assume that
we do know how to construct D-branes on the covering space
and ask how to construct D-branes on the orbifold, i.e. the
GSO-projected theory.
The orbifold action on states of
charge $q$ is given by 
\beq\label{orbac}
|q\rangle \to e^{2\pi i q} |q\rangle\ \ \ .
\eeq
In this way, we project on a discrete set of charges.
In a geometrical context, the charges can be
interpreted as momenta, and we are effectively compactifying
on a circle.
To describe D-branes in that
situation, we can make use of methods developed in
\cite{wati}. 
\vspace{1cm}
\fig{Pre-images of branes related by the spectral flow operator}
{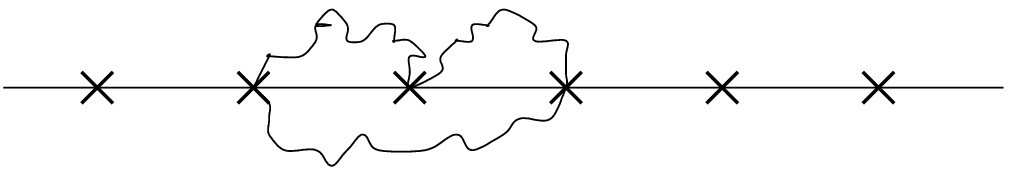}{10truecm}
\figlabel{\cover}
\vspace{1cm}
Here, the D-branes on the circle are described by images of
the translation operator along the real line. In our
situation, this translation operator is given by spectral
flow. To describe D-branes, we have to sum over images
of D-branes in the non-GSO projected theory under
spectral flow. 
\smallskip

Moving to the Gepner point of the compactification means
that we enhance the chiral algebra by so many operators
that we only have a finite number of primary fields.
Most of the branes would not preserve the full symmetry
group. Those branes which do preserve the enhanced symmetry
have only finitely many
images under  the spectral flow operator. In this case,
the spectral flow generates a finite group $\Gamma$, and
we can directly apply the methods developed in Section 3.
The rationality of the model enables us to describe the
D-branes on the covering space as boundary states
and one can 
find a large set of boundary states \cite{RS} which 
respect the $\CN=2$ world-sheet algebras of each minimal 
model factor of the Gepner model separately. 
\medskip

To this end we start 
with Cardy boundary states of the tensor product theory. They  
are given by the expression that involves the S-matrices of 
minimal models and the $SO(2)$ theory (see e.g.\ \cite{RS}). 
Cardy's boundary states belong  to some gluing condition $W(z) 
= \Omega \bar W(\bar z), z = \bar z,$ which becomes $J_{i} (z) 
= - \bar J_i(\bar z)$ on the $U(1)$-currents of the individual 
theories. This means that they are A-type boundary conditions 
in the sense of \cite{OoOzYi}.  
\smallskip

The boundary states $|I\rangle =: |\Lambda, \Xi\rangle$ we have just 
described depend on a spin vector $\Lambda = (L_1, \dots, L_r)$ and 
a charge vector $\Xi = (S_0;M_1, \dots, M_r; S_1, \dots, S_r)$.
{}From these states in the tensor product theory we can pass to 
boundary states of the Gepner model using the general strategy 
explained above. The projected boundary states in the 
orbifold theory are given by 
\beq\label{geporb}
 |\Lambda,\Xi\rangle_\proj \ = \  \frac{1}{\sqrt{K 2^r}} 
   \sum_{\nu,\nu_i} \, (-1)^\nu (-1)^{\frac{\hat s_0^2}{2}} \, \, 
   |\Lambda, \Xi 
   + \nu\beta_0+\nu_1\beta_1+\ldots +\nu_r\beta_r\rangle\ \ . 
\eeq 
Here, the element $\hat s_0$ is an operator acting on closed string 
states which measures the value $s_0$. The whole factor $(-1)^{\hat 
s_0^2/2}$ is needed to guarantee that in the open string partition 
function (similar to the closed string partition function) fields 
are counted with a phase factor referring to their ghost charge. 
Making use of Cardy's formalism we obtain the expressions
established in 
\cite{RS},  
\beq
\label{rsstate}
       \kket{\alpha} \ := \  \kket{\Lambda,\Xi}_\proj\ =\ 
         \sum_{\mathbf \lambda,\mathbf \mu; 
      \tilde Q (\mathbf \lambda,\mathbf \mu) = 0} (-1)^{\frac{s_0^2}{2}} \ \
        B^{\lambda,\mu}_\alpha\ \kket{\lambda,\mu}\ .
\eeq
with the coefficients:
\beq
\label{rscoeff}
        B^{\lambda,\mu}_{\alpha}\ =\ {\sqrt{K 2^r}\over 2}
         \, e^{-i\pi \frac{s_0 S_0}{2}}\ 
        \prod_{j=1}^r{1\over{\sqrt{\sqrt{2}(k_j+2)}}}
                {{\sin(l_j,L_j)_{k_j}}\over
        {\sqrt{\sin(l_j,0)_{k_j}}}}\ e^{i\pi{{m_j M_j}\over{k_j+2}}}
        \ e^{-i\pi{{s_j S_j}\over{2}}}\ .
\eeq
Here $(l,l')_k = \pi (l+1)(l'+1)/(k+2)$. For these A-type boundary 
states the Ishibashi states are built on diagonal primary states, 
i.e.\ states in the untwisted sector, in accordance with our general 
theory in Section 2. The associated partition functions  
(\ref{Zproj}) acquire the following form (see also \cite{RS}):
\beqa
\label{apart}
        Z_{\tilde \alpha \alpha}^A(q)& =& {1\over 2}\ \sum_{\lambda',\mu'}
        \sum_{\nu=0}^{K-1}\sum_{\nu_i=0,1}\ (-1)^{s_0' + S_0 - \tilde S_0} 
        \ \delta^{(4)} _{s_0' + \tilde S_0 - S_0 + \nu + 2 \sum\nu_i -2 }
        \nn \\[3mm] & & \hspace*{1cm} \times \   
        \prod_{j=1}^rN^{L_j}_{l_j',\tilde L_j}
        \ \delta^{(2k_j+4)}_{\nu - M_j+ \tilde
        M_j+m_j'}\ \delta^{(4)}_{s_j'+\tilde S_j - S_j + \nu + 2\nu_j}
        \  \chi^{\lambda'}_{\mu'}(q)\ . 
\eeqa
The factor $1/2$ in front of the right hand side accounts for the fact 
that field identification causes each character to appear twice when 
we sum over $\lambda',\mu'$ taken from the extended range.  
\smallskip

There is one important difference between the orbits in equation 
(\ref{geporb}) and the summation over D0-brane states on the real 
line. While the latter are infinite (due to the infinite extension
of the real line), the former have finite length. In particular, 
there can be orbits of different length in a Gepner model. The
identity orbit has length K, but in some cases there can be
orbits of length $K/2$. This means that the boundary state is
invariant under an application of the spectral flow to the
power $K/2$. This situation can be treated similarly to the
case that there is a fix-point under a geometrical $\BZ_2$. 
The boundary states get labeled by an orbit label and a 
representation of the $\BZ_2$, which is a sign. They cannot
be directly obtained from the covering space. However, the
sum of an orbit labeled ``$+$'' and an orbit labeled ``$-$''
has again an interpretation on the covering space. (In the
language of \cite{DM} the resulting brane corresponds
to choosing the regular representation.) This has been 
worked out in \cite{BruSch} (see also \cite{FSW}). In the 
rest of this paper, we will restrict  ourselves to the case of 
branes that can be obtained as invariant objects that
come down from the covering space. That means that we are
choosing the regular representation in the case that
there is a non-trivial stabilizer.
\smallskip

If the two boundary conditions $\a,\tilde \a$ appearing in eq.\ (\ref{apart})
are both labeled by monodromy invariant orbits, they give rise to a monodromy 
invariant open string spectrum, i.e.\ to a spectrum that contains only 
odd-integer charges. One should recall, however, that non-invariant 
orbits of $\Gamma$ are also admissible as labels for boundary conditions. 
The condition for a supersymmetric open string spectrum consisting of 
monodromy invariant states is that the U(1) charge of the two orbits 
labeling the boundary conditions $\a$ and $\tilde a$ differs by an even 
integer.

\subsection{The boundary OPE in Gepner models}

We have seen that there is a set of Gepner model boundary states 
(\ref{geporb}) which can be understood as invariant linear 
combinations of boundary states on the covering space as in 
our general discussion in Subsection 3.2. Each boundary 
operator $\phi^{[\Lambda , \Xi], [\Lambda', \Xi']}_{\lambda, 
\mu}$,  where $[\Lambda ,\Xi ], [\Lambda', \Xi']$ are orbit 
labels, can be understood as coming from an operator on the 
covering space, i.e.\ we can identify 
\beq \label{opcop}  
\Psi^{[\Lambda \Xi], [\Lambda' \Xi']}_{\lambda,\mu} \ := \ 
\Psi^{(\Lambda ,\Xi+\nu_0 \beta_0+ \nu_i \beta_i)
(\Lambda', \Xi'+\nu_0'\beta_0 + \nu_r'\beta_r)}_{\lambda, \mu}
\eeq
Here, $(\Lambda, \Xi +\nu_0 \beta_0 +\nu_i\beta_i)$ and 
$(\Lambda', \Xi' +\nu_0' \beta_0 +\nu_i'\beta_i)$ label branes on 
the covering space such that the field $(\lambda,\mu)$ propagates 
between them. 
\smallskip

When we want to multiply boundary operators in the Gepner model, 
we can use the identification (\ref{opcop}) and then interpret
the right hand side as a product of boundary operators in the
individual models, i.e.\ 
\beq \label{Gepbo}  
\Psi^{(\Lambda,\Xi)(\Lambda',\Xi')}_{\lambda,\mu}(x) \ := \ 
\psi^{S_0 S'_0}_{s_0\, ;\, 0} (x) \ 
 \prod_{i=1}^r \ \psi^{(L_iM_iS_i)(L'_iM'_iS'_i)}_{(l_im_is_i)\, ;\, i}(x)    
\eeq
where $\Lambda = (L_1, \dots, L_r), \Xi = (S_0;M_1, \dots,M_r,S_1, 
\dots,S_r)$ etc.\ and $\psi^{\dots}_{.\, ;\, i}, i=1, \dots,r,$ denote 
boundary operators within the $i^{th}$ minimal model. These operators 
are multiplied using eqs.\ (\ref{MMope}) with the appropriate structure
constants depending on the level $k_i$ of the $i^{th}$ minimal model. 
\smallskip

In eq.\ (\ref{Gepbo}) there appears one more set of boundary operators, 
namely the operators $\psi^{\dots}_{.\, ;\, 0}$ which come with the external 
fermions. They are multiplied according to 
\beq \label{Fope} 
\psi^{S_0,S'_0}_{s_0\, ;\, 0}(x_1) \ \psi^{S'_0, S''_0}_{s'_0\, ;\, 0} (x_2) \ = \ 
 (-1)^{\frac{s'_0}{8}(S_0 + s_0 - S_0')} \ \delta_{S_0',s_0  \hp S_0} \, 
 \delta_{S''_0,s'_0 \hp S'_0} \, \delta_{S''_0, s''_0 \hp S_0} \ 
 \psi^{S_0,S''_0}_{s''_0\, ;\, 0}(x_2)     
\eeq
for $x_1 < x_2$, as usual. The rules (\ref{opcop},\ref{Gepbo}) along 
with the expansions (\ref{MMope}) and (\ref{Fope}) allow to compute 
the operator products of arbitrary boundary operators for untwisted
A-type branes in Gepner models. 
\smallskip

According to Subsection 3.3, the results 
one obtains from these four equations do not depend on the choice of the 
representative $(\Lambda,\Xi +\nu_0 \beta_0 +\nu_i\beta_i) \in [\Lambda,\Xi]$ 
that one has to make in eq.\ (\ref{opcop}). Geometrically, the freedom 
is associated with the choice of a particular brane on the cover 
that is used to represent the brane $[\Lambda,\Xi]$ in the orbifold.   
If some open string operator $(\lambda,\mu)$ propagates between two 
pre-images $(\Lambda, \Xi),(\Lambda',\Xi')$ on the covering space, 
then is also propagates between two pre-images $(\Lambda, \Xi+\nu_0 
\beta_0 + \nu_r\beta_r), (\Lambda',\Xi'+\nu_0\beta_0+\nu_r\beta_r)$ 
which are obtained from the first two by applying the simple current
$\nu_0\beta_0+\nu_r\beta_r$. In  Figure 2  this is reflected by the 
fact that the starting point of a string can be moved from one brane 
to the next, if the endpoint is moved accordingly. 

\section{Superpotential for A-type branes on the quintic}

We are finally able to combine the discussion of Section 2 with 
the explicit formulas (\ref{opcop},\ref{Gepbo},\ref{MMope},
\ref{Fope}) for computations in the internal CFT to calculate
terms in the superpotential of untwisted A-type branes in Gepner
models. We shall illustrate these calculations through one 
example of an A-type brane in the quintic. 
\medskip
 
In the example we shall discuss, the boundary state and its 
geometrical interpretation is known. A set of special Lagrangian
submanifolds on the quintic given by Im $\omega_jz_j$ with
$\omega_j^5=1$ was described in \cite{BDLR}. Topologically, these 
submanifolds are the real projective space $\QR\QP^3$. In particular, 
$\pi_1(\QR\QP^3)=\QZ_2$ and therefore there are no continuous moduli 
in the geometric (large volume) regime. On the other hand, one can 
identify a set of boundary states which is believed to correspond to 
the geometric cycles. This was checked \cite{BDLR} by comparing the 
intersection numbers which can be computed both for the geometric 
cycles and the boundary states. It was observed in \cite{BDLR} that 
there exits one massless mode in the boundary CFT description of these
cycles. We will show below that this massless operator is not truly 
marginal, or, in other words, that there is a superpotential term 
for this operator. 
\smallskip

The Gepner model for the quintic is the model $(k=3)^5$ and the boundary 
states corresponding to the cycles we consider are those for which $L_i=1$, 
where $i=1,\dots ,5$. In this case, there appears a single marginal operator
in the spectrum whose internal part  is given by 
\beq
\psi (x) \ = \ \Psi^{[{\bf 1}, \Xi][{\bf 1},\Xi]}_{(1,1,0)^{\times_i}}(x)
\eeq
where ${\bf 1}$ denotes the vector $\Lambda = (1,1,1,1,1) = {\bf 1}$
and we assign the same label $(1,1,0)$ to all the minimal models. 
We shall also need the field that is obtained from $\Psi$ by applying 
one unit of spectral flow, corresponding to the auxiliary $F$-field in 
the same multiplet. In the internal sector, this amounts to a fusion 
with the anti-chiral field of highest charge. The resulting internal 
part of the auxiliary field is
\beq
\psi_F (x) \ = \ \Psi^{[{\bf 1}, \Xi][{\bf 1},\Xi]}_{(2,-2,0)^{\times_i}}  
(x) \ \ . \eeq
The external and superghost contributions to the full vertex operators 
were described in Section 2. Let us also recall from there that the 
correlation function has to include three conformal ghost $c$.
\smallskip

Because of the $SL(2,\QR)$ invariance of the theory, the ghost 
contribution cancels the dependence on world-sheet coordinates in 
a 3-point function. The value of the correlator can be determined 
by successive OPEs of the vertex operators discussed in Section 2.
Since the external part of the scalar is the identity field, the 
full amplitude is essentially a product of OPE coefficients
of the fields in the internal sector.

According to our formulas in Section 5.3, these OPE coefficients 
can be obtained from the OPE on the `covering space'. Therefore 
we just have to choose pre-images on the covering space, such 
that the fields with subscripts $(1,1,0)$ and $(2,-2,0)$ 
propagate. We start out by multiplying $\psi$ with itself. 
The two operators in the theory for the covering space are 
given by 
\beq\label{schritteins}
 \prod_i \psi_{(1,1,0)}^{(1, M_i-2, S_i)(2, M_i-1, S_i)} 
\ \ \ \mbox{ and } \ \ \  \prod_i \psi_{(1,1,0)}^{(2, M_i-1, S_i)
  (1,M_i,S_i)}\ \ . 
\eeq
Note that we have applied a field identification on two of the 
superscripts labeling boundary conditions to replace $L=1$ by $L=2$.
When we calculate the OPE of the two fields in (\ref{schritteins}), 
there appear several contributions out of which only one term can 
contribute to a 2-point function with $\psi_F$. This term is 
proportional to the field 
\beq \label{Fdag} 
      \prod_i \psi_{(2,2,0)}^{(1,M_i-2,S_i)(1,M_i,S_i)}\ \ .  
\eeq
In fact, if we choose to represent $\psi_F$ on the covering space 
by a field of the form 
\beq\label{schrittzwei}
  \prod_i \psi_{(2,-2,0)}^{(1,M_i,S_i)(1,M_i-2,S_i)}\   
\eeq
then the OPE between the operators (\ref{Fdag}) and (\ref{schrittzwei}) 
gets a contribution from the identity field. From the OPE of the two 
operators in (\ref{schritteins}) and a consecutive OPE of the fields 
(\ref{Fdag}), (\ref{schrittzwei}) we obtain the following coefficient
of the correlation function 
\beq
C_3 \ = \ \left( \Fus{(2,M-1,S)}{(2,2,0)}{(1,1,0)}{(1,1,0)}{(1,M-2,S)}
{(1,M,S)} \Fus{(1,M-2,S)}{(0,0,0)}{(2,-2,0)}{(2,2,0)}{(1,M,S)}
{(1,M,S)} \CG{2}{2}{0}{-2}{2}{0} \right)^5
\eeq
The $5^{th}$ power comes from the five identical factors that 
contribute to our correlator. We can decompose the fusing matrices 
further into WZW fusing matrices and phase factors coming from 
the two $U(1)$ theories. Actually, the latter do not appear in 
this special case so that we obtain  
$$
C_3 \ =\ \left( \Fus{2}{2}{1}{1}{1}{1} \Fus{1}{0}{2}{2}{1}{1} 
         \CG{2}{2}{0}{-2}{2}{0} \right)^5
     \ \ . 
$$
Here, $F$ and $[.]$ denote the fusing matrix of $\su$ and the 
Clebsch-Gordan coefficients of $su(2)$, respectively. To obtain 
the full correlation function in the Gepner model, we finally
have to take into account the expectation value of the identity
with boundary conditions $({\bf 1}, \Xi) $. It is given by
\beq
\vev{{\bf 1}}_{({\bf 1}, \Xi)}\ = \ 
\left( \frac{S_{1}^{\ 0}}{S_{0}^{\ 0}} \right)^5 \ . 
\eeq
Here, the matrix elements $S_{l}^{\ 0}$ are obtained directly 
{}from the S-matrix of the $\su$ theory since the phase factors
{}from the $U(1)$ theory drop out. The resulting expression for 
the correlator with one particular ordering of operators is
\beq
 \left( \Fus{2}{2}{1}{1}{1}{1} \Fus{1}{0}{2}{2}{1}{1} 
\CG{2}{2}{0}{-2}{2}{0}\right)^5
\left( \frac{S_{1}^{\ 0}}{S_{0}^{\ 0}} \right)^5\ 
\ \ . 
\eeq
This result is symmetric with respect to exchanging the last 
two fields. Hence, summation over inequivalent orderings of 
the three insertion points simply produces a factor $2$. In 
conclusion, we have shown that the relevant cubic term in 
the superpotential of our brane is given by 
\beq
\sim \ 2 \left( \Fus{2}{2}{1}{1}{1}{1} \Fus{1}{0}{2}{2}{1}{1} 
\CG{2}{2}{0}{-2}{2}{0}\right)^5
\left( \frac{S_{1}^{\ 0}}{S_{0}^{\ 0}} \right)^5\ F\, \phi\, \phi
\ \ . 
\eeq
{}From the fact that our result 
does not vanish, we conclude that in first order perturbation 
theory, the modulus generated by $\Psi$ gets lifted so that 
there is no associated flat direction in the world-volume 
theory.

\section{Conclusions and outlook}

In this paper we have computed the boundary operator product 
expansions for untwisted (or {\em projected}) A-type boundary 
states in Gepner models. The main idea was to compute the 
expansions on a `covering space', i.e.\ for products of 
minimal models, first and then to (GSO) project them to the 
Gepner model. We applied our results to one particular 
example in which we computed the cubic term of the 
superpotential for a massless field. 
\smallskip

There are a number of possible extension. First of all, we 
are not restricted to the evaluation of 3-point functions
or, equivalently, third order terms of the superpotential. 
As is well known in conformal field theory, all correlation 
functions can be recovered from the operator product 
expansions with the help of Ward identities. In this sense, 
we have completely solved the boundary correlators by 
giving expressions for the coefficients of the OPE. 
Of course, the computations may still be rather involved. 
\smallskip

Furthermore, we can compute correlators for fields of 
arbitrary masses, including tachyonic fields. This makes 
it possible to study bound states of unstable brane 
configurations in Gepner models. From a CFT point of
view, one has to perturb the open string action by a 
(marginally) relevant operator. Knowledge of the 
boundary OPE enables us to compute the $\beta$-
function along the RG trajectory generated by this
perturbation. The bound state appears at a point where 
the $\beta$-function vanishes, giving rise to a new 
conformal field theory. Since the structure constants 
in our Gepner model correlators are essentially given
by data of the $\su$ WZW model, the results of 
\cite{ARS} on bound states in WZW models may partly 
be carried over to an $\CN=2$ minimal model (see also
\cite{RRS} for an analysis of related problems in 
Virasoro minimal models).  This suggests that arbitrary 
branes in Gepner models can be obtained as bound states 
of branes with $L=0$.   
\smallskip

Similar phenomena occur for B-type boundary conditions
in Gepner models where all the known boundary theories
appear as bound states of fractional branes, as was 
argued in \cite{DouDia}. The work by Douglas and 
Diaconescu contains terms of the superpotential 
for B-type branes. It would be interesting to verify 
their proposal through exact world-sheet computations. 
The main part of our analysis, namely Sections 2-4, 
provide a solid basis for extending our constructions
so that they incorporate projected B-type boundary 
states.  
\smallskip

Throughout this work we focused our attention on 
boundary correlators and omitted all questions related 
to bulk-boundary couplings. In the context of Gepner 
models it would be of particular interest to compute 
couplings of boundary fields to the bulk fields which 
generate deformations of the complex- and K\"ahler 
structure. Such couplings encode the dependence of 
brane moduli spaces on the K\"ahler/complex structure.
A better understanding of these couplings could 
provide more insight into the decoupling conjecture 
of \cite{BDLR}. We plan to come back to some of these 
issues in the future. 
\bigskip
\bigskip

\noindent
{\bf Acknowledgements:} We would like to thank E.\ Diaconescu, 
J.\ Distler, M.\ Douglas, 
S.\ Fredenhagen, J.\ Fuchs, A.\ Hanany, A.\ Iqbal,
S.\ Kachru, A.\ Karch, A.\ Lawrence, A.\ Recknagel, Chr.\ R\"omelsberger,
C.\ Schweigert 
and E.\ Silverstein. One of us (V.S.) is grateful to the string groups at 
Rutgers University and at the IAS Princeton for their hospitality. 
I.B. thanks the Erwin Schr\"odinger Institut (ESI).

\newpage
\bibliography{supercft}

\begingroup\raggedright\begin{thebibliography}{10}

\bibitem{wittph}
E.~Witten, ``Phases of {N = 2} theories in two dimensions,'' {\em Nucl. Phys.}
  {\bf B403} (1993) 159--222,
  \href{http://xxx.lanl.gov/abs/hep-th/9301042}{{\tt hep-th/9301042}}.

\bibitem{Gepn1}
D.~Gepner, ``Exactly solvable string compactifications on manifolds of {SU(N)}
  holonomy,'' {\em Phys. Lett.} {\bf B199} (1987) 380--388.

\bibitem{Gepn2}
D.~Gepner, ``Space-time supersymmetry in compactified string theory and
  superconformal models,'' {\em Nucl. Phys.} {\bf B296} (1988) 757.

\bibitem{BDLR}
I.~Brunner, M.~R. Douglas, A.~Lawrence, and C.~{R\"omelsberger}, ``D-branes on
  the quintic,'' \href{http://xxx.lanl.gov/abs/hep-th/9906200}{{\tt
  hep-th/9906200}}.

\bibitem{RS}
A.~Recknagel and V.~Schomerus, ``D-branes in {Gepner} models,'' {\em Nucl.
  Phys.} {\bf B531} (1998) 185,
  \href{http://xxx.lanl.gov/abs/hep-th/9712186}{{\tt hep-th/9712186}}.

\bibitem{DiaRom}
D.-E. Diaconescu and C.~{R\"omelsberger}, ``D-branes and bundles on elliptic
  fibrations,'' \href{http://xxx.lanl.gov/abs/hep-th/9910172}{{\tt
  hep-th/9910172}}.

\bibitem{KLLW}
P.~Kaste, W.~Lerche, C.~A. Lutken, and J.~Walcher, ``D-branes on
  {K3}-fibrations,'' \href{http://xxx.lanl.gov/abs/hep-th/9912147}{{\tt
  hep-th/9912147}}.

\bibitem{ej}
D.-E. Diaconescu and J.~Gomis, ``Fractional branes and boundary states in
  orbifold theories,'' \href{http://xxx.lanl.gov/abs/hep-th/9906242}{{\tt
  hep-th/9906242}}.

\bibitem{Scheid}
E.~Scheidegger, ``D-branes on some one- and two-parameter {Calabi-Yau}
  hypersurfaces,'' \href{http://xxx.lanl.gov/abs/hep-th/9912188}{{\tt
  hep-th/9912188}}.

\bibitem{DFR2}
M.~R. Douglas, B.~Fiol, and C.~{R\"omelsberger}, ``The spectrum of {BPS} branes
  on a noncompact {Calabi-Yau},''
  \href{http://xxx.lanl.gov/abs/hep-th/0003263}{{\tt hep-th/0003263}}.

\bibitem{DFR1}
M.~R. Douglas, B.~Fiol, and C.~{R\"omelsberger}, ``Stability and {BPS}
  branes,'' \href{http://xxx.lanl.gov/abs/hep-th/0002037}{{\tt
  hep-th/0002037}}.

\bibitem{DouDia}
D.-E. Diaconescu and M.~R. Douglas, ``D-branes on stringy {Calabi-Yau}
  manifolds,'' \href{http://xxx.lanl.gov/abs/hep-th/0006224}{{\tt
  hep-th/0006224}}.

\bibitem{HIV}
K.~Hori, A.~Iqbal, and C.~Vafa, ``D-branes and mirror symmetry,''
  \href{http://xxx.lanl.gov/abs/hep-th/0005247}{{\tt hep-th/0005247}}.

\bibitem{suresh1}
S.~Govindarajan and T.~Jayaraman, ``On the {Landau-Ginzburg} description of
  boundary {CFTs} and special {Lagrangian} submanifolds,''
  \href{http://xxx.lanl.gov/abs/hep-th/0003242}{{\tt hep-th/0003242}}.

\bibitem{suresh2}
S.~Govindarajan, T.~Jayaraman, and T.~Sarkar, ``Worldsheet approaches to
  {D-branes} on supersymmetric cycles,''
  \href{http://xxx.lanl.gov/abs/hep-th/9907131}{{\tt hep-th/9907131}}.

\bibitem{GJSIII}
S.~Govindarajan, T.~Jayaraman, and T.~Sarkar, ``On {D-branes} from gauged
  linear sigma models,'' \href{http://xxx.lanl.gov/abs/hep-th/0007075}{{\tt
  hep-th/0007075}}.

\bibitem{NakNoz}
M.~Naka and M.~Nozaki, ``Boundary states in {G}epner models,''
  \href{http://xxx.lanl.gov/abs/hep-th/0001037}{{\tt hep-th/0001037}}.

\bibitem{BruSch}
I.~Brunner and V.~Schomerus, ``D-branes at singular curves of {Calabi-Yau}
  compactifications,'' \href{http://xxx.lanl.gov/abs/hep-th/0001132}{{\tt
  hep-th/0001132}}.

\bibitem{FSW}
J.~Fuchs, C.~Schweigert, and J.~Walcher, ``Projections in string theory and
  boundary states for {Gepner} models,''
  \href{http://xxx.lanl.gov/abs/hep-th/0003298}{{\tt hep-th/0003298}}.

\bibitem{KKLM1}
S.~Kachru, S.~Katz, A.~Lawrence, and J.~McGreevy, ``Open string instantons and
  superpotentials,'' \href{http://xxx.lanl.gov/abs/hep-th/9912151}{{\tt
  hep-th/9912151}}.

\bibitem{KKLM2}
S.~Kachru, S.~Katz, A.~Lawrence, and J.~McGreevy, ``Mirror symmetry for open
  strings,'' \href{http://xxx.lanl.gov/abs/hep-th/0006047}{{\tt
  hep-th/0006047}}.

\bibitem{RS2}
A.~Recknagel and V.~Schomerus, ``Boundary deformation theory and moduli spaces
  of {D-branes},'' {\em Nucl. Phys.} {\bf B545} (1999) 233,
  \href{http://xxx.lanl.gov/abs/hep-th/9811237}{{\tt hep-th/9811237}}.

\bibitem{Sen}
A.~Sen, ``Tachyon condensation on the brane antibrane system,'' {\em JHEP} {\bf
  08} (1998) 012, \href{http://xxx.lanl.gov/abs/hep-th/9805170}{{\tt
  hep-th/9805170}}.

\bibitem{HKM}
J.~A. Harvey, D.~Kutasov, and E.~J. Martinec, ``On the relevance of tachyons,''
  \href{http://xxx.lanl.gov/abs/hep-th/0003101}{{\tt hep-th/0003101}}.

\bibitem{ARS}
A.~Y. Alekseev, A.~Recknagel, and V.~Schomerus, ``Brane dynamics in background
  fluxes and non-commutative geometry,'' {\em JHEP} {\bf 05} (2000) 010,
  \href{http://xxx.lanl.gov/abs/hep-th/0003187}{{\tt hep-th/0003187}}.

\bibitem{Sen2}
J.~Majumder and A.~Sen, ``Non-{BPS} {D}-branes on a {Calabi}-{Yau} orbifold,''
  \href{http://xxx.lanl.gov/abs/hep-th/0007158}{{\tt hep-th/0007158}}.

\bibitem{DoHu}
M.~R. Douglas and C.~Hull, ``D-branes and the noncommutative torus,'' {\em
  JHEP} {\bf 02} (1998) 008, \href{http://xxx.lanl.gov/abs/hep-th/9711165}{{\tt
  hep-th/9711165}}.

\bibitem{ChuHo}
C.-S. Chu and P.-M. Ho, ``Noncommutative open string and {D}-brane,'' {\em
  Nucl. Phys.} {\bf B550} (1999) 151,
  \href{http://xxx.lanl.gov/abs/hep-th/9812219}{{\tt hep-th/9812219}}.

\bibitem{Scho}
V.~Schomerus, ``D-branes and deformation quantization,'' {\em JHEP} {\bf 06}
  (1999) 030, \href{http://xxx.lanl.gov/abs/hep-th/9903205}{{\tt
  hep-th/9903205}}.

\bibitem{SeiWit}
N.~Seiberg and E.~Witten, ``String theory and noncommutative geometry,'' {\em
  JHEP} {\bf 09} (1999) 032, \href{http://xxx.lanl.gov/abs/hep-th/9908142}{{\tt
  hep-th/9908142}}.

\bibitem{MooSei}
G.~Moore and N.~Seiberg, ``Classical and quantum conformal field theory,'' {\em
  Commun. Math. Phys.} {\bf 123} (1989) 177.

\bibitem{Run}
I.~Runkel, ``Boundary structure constants for the {A-series Virasoro} minimal
  models,'' {\em Nucl. Phys.} {\bf B549} (1999) 563,
  \href{http://xxx.lanl.gov/abs/hep-th/9811178}{{\tt hep-th/9811178}}.

\bibitem{FFFS1}
G.~Felder, J.~Frohlich, J.~Fuchs, and C.~Schweigert, ``Correlation functions
  and boundary conditions in {RCFT} and three-dimensional topology,''
  \href{http://xxx.lanl.gov/abs/hep-th/9912239}{{\tt hep-th/9912239}}.

\bibitem{FFFS2}
G.~Felder, J.~Frohlich, J.~Fuchs, and C.~Schweigert, ``Conformal boundary
  conditions and three-dimensional topological field theory,'' {\em Phys. Rev.
  Lett.} {\bf 84} (2000) 1659,
  \href{http://xxx.lanl.gov/abs/hep-th/9909140}{{\tt hep-th/9909140}}.

\bibitem{zuber}
R.~E. Behrend, P.~A. Pearce, V.~B. Petkova, and J.-B. Zuber, ``Boundary
  conditions in rational conformal field theories,'' {\em Nucl. Phys.} {\bf
  B570} (2000) 525, \href{http://xxx.lanl.gov/abs/hep-th/9908036}{{\tt
  hep-th/9908036}}.

\bibitem{Run2}
I.~Runkel, ``Structure constants for the {D-series Virasoro} minimal models,''
  {\em Nucl. Phys.} {\bf B579} (2000) 561,
  \href{http://xxx.lanl.gov/abs/hep-th/9908046}{{\tt hep-th/9908046}}.

\bibitem{FucSchI}
J.~Fuchs and C.~Schweigert, ``Symmetry breaking boundaries. {I}: {General}
  theory,'' {\em Nucl. Phys.} {\bf B558} (1999) 419,
  \href{http://xxx.lanl.gov/abs/hep-th/9902132}{{\tt hep-th/9902132}}.

\bibitem{FucSchII}
J.~Fuchs and C.~Schweigert, ``Symmetry breaking boundaries. {II}: {More}
  structures, examples,'' {\em Nucl. Phys.} {\bf B568} (2000) 543,
  \href{http://xxx.lanl.gov/abs/hep-th/9908025}{{\tt hep-th/9908025}}.

\bibitem{Lew}
D.~C. Lewellen, ``Sewing constraints for conformal field theories on surfaces
  with boundaries,'' {\em Nucl. Phys.} {\bf B372} (1992) 654--682.

\bibitem{PrSaSt}
G.~Pradisi, A.~Sagnotti, and Y.~S. Stanev, ``Completeness conditions for
  boundary operators in 2d conformal field theory,'' {\em Phys. Lett.} {\bf
  B381} (1996) 97--104, \href{http://xxx.lanl.gov/abs/hep-th/9603097}{{\tt
  hep-th/9603097}}.

\bibitem{multifsI}
J.~Fuchs, B.~Schellekens, and C.~Schweigert, ``Twining characters, orbit lie
  algebras, and fixed point resolution,''
  \href{http://xxx.lanl.gov/abs/q-alg/9511026}{{\tt q-alg/9511026}}.

\bibitem{multifsII}
J.~Fuchs, B.~Schellekens, and C.~Schweigert, ``The resolution of field
  identification fixed points in diagonal coset theories,'' {\em Nucl. Phys.}
  {\bf B461} (1996) 371--406,
  \href{http://xxx.lanl.gov/abs/hep-th/9509105}{{\tt hep-th/9509105}}.

\bibitem{Gree}
B.~R. Greene, ``String theory on {Calabi-Yau} manifolds,''
  \href{http://xxx.lanl.gov/abs/hep-th/9702155}{{\tt hep-th/9702155}}.

\bibitem{DM}
M.~R. Douglas and G.~Moore, ``D-branes, quivers, and {ALE} instantons,''
  \href{http://xxx.lanl.gov/abs/hep-th/9603167}{{\tt hep-th/9603167}}.

\bibitem{wati}
W.~Taylor, ``D-brane field theory on compact spaces,'' {\em Phys. Lett.} {\bf
  B394} (1997) 283--287, \href{http://xxx.lanl.gov/abs/hep-th/9611042}{{\tt
  hep-th/9611042}}.

\bibitem{OoOzYi}
H.~Ooguri, Y.~Oz, and Z.~Yin, ``D-branes on {Calabi-Yau} spaces and their
  mirrors,'' {\em Nucl. Phys.} {\bf B477} (1996) 407--430,
  \href{http://xxx.lanl.gov/abs/hep-th/9606112}{{\tt hep-th/9606112}}.

\bibitem{RRS}
A.~Recknagel, D.~Roggenkamp, and V.~Schomerus, ``On relevant boundary
  perturbations of unitary minimal models,''
  \href{http://xxx.lanl.gov/abs/hep-th/0003110}{{\tt hep-th/0003110}}.

\end{thebibliography}\endgroup
\bibliographystyle{utphys}

\end{document}